\documentclass[bibyear]{aa}

\usepackage{graphicx}

\usepackage{txfonts}

\usepackage{natbib}
\bibpunct{(}{)}{;}{a}{}{,}
\usepackage{booktabs}
\usepackage{multicol}
\usepackage{graphicx}
\usepackage{fancyhdr}

\usepackage{amsmath}	
\usepackage{amssymb}	
\usepackage[inter-unit-product=\cdot]{siunitx}
\usepackage{enumerate}
\usepackage{multirow}

\usepackage{mathtools}

\usepackage{longtable}
\usepackage{tabularx}
\usepackage{xcolor} 
\usepackage{ulem} 
\usepackage{comment}
\usepackage{cuted}

\usepackage{soul}

\usepackage{placeins}
\usepackage{stfloats}
\usepackage{float}

\usepackage{hyperref}
\hypersetup{colorlinks=true,urlcolor=black, citecolor=blue, linkcolor=black}

\makeatletter
\renewcommand*\aa@pageof{, page \thepage{} of \pageref*{LastPage}}
\makeatother

\newcommand{\orcidicon}[1]{\href{https://orcid.org/#1}{\includegraphics[width=11pt]{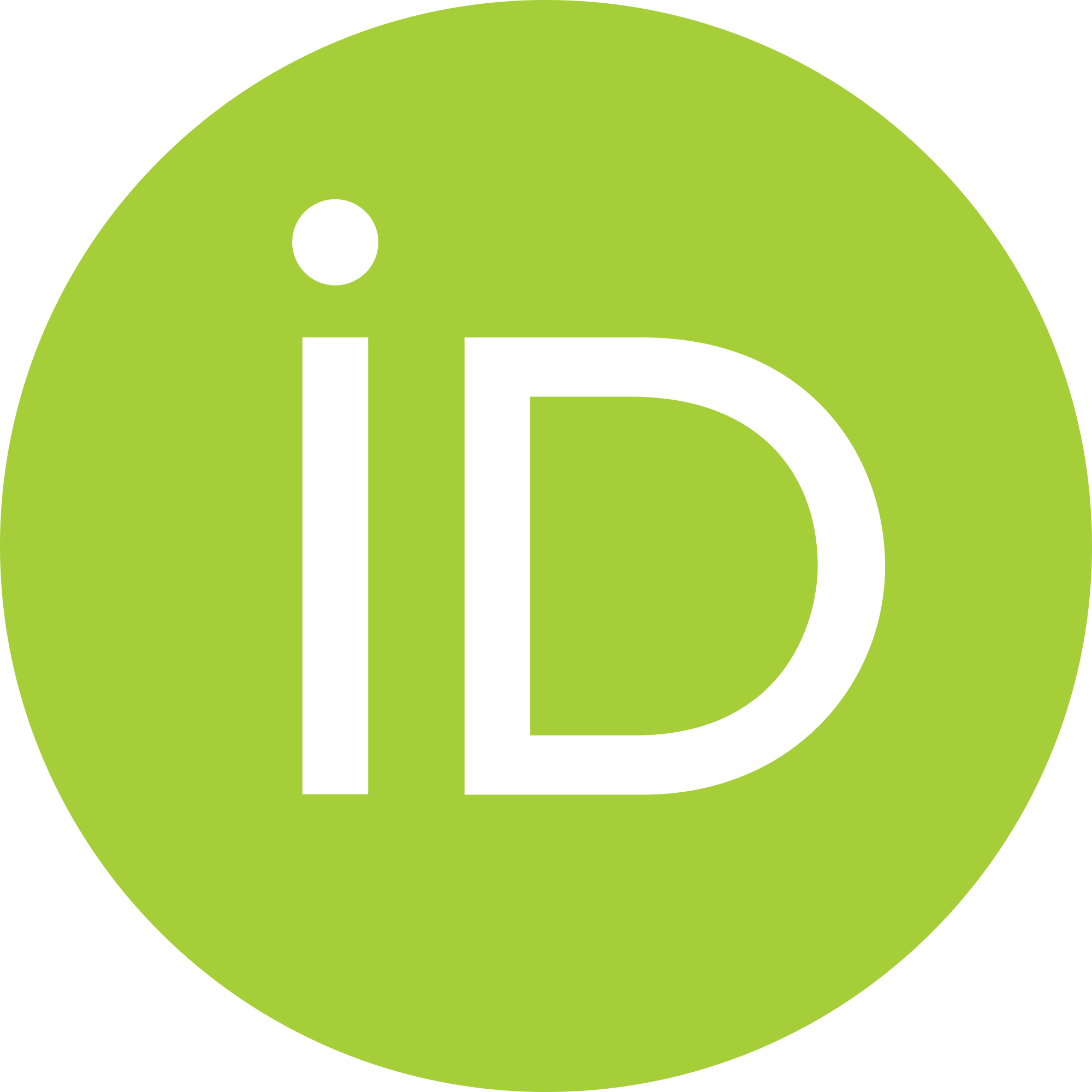}}}
\newcommand{\orcid}[1]{\href{https://orcid.org/#1}{\protect\orcidicon{#1}}}

\definecolor{steelblue}{rgb}{0.274 0.510 0.706}

\begin{document}

   \title{Can current models predict the local black hole merger rate?}
    \titlerunning{}

   \author{
    Lumen Boco\inst{1}
    \orcid{0000-0003-3127-922X} \thanks{\href{mailto:lumen.boco@uni-heidelberg.de}{lumen.boco@uni-heidelberg.de}},
    Michele Bosi\inst{2,3}
    \orcid{0009-0000-8215-6698},
    Cecilia Sgalletta\inst{1}
    \orcid{0009-0003-7951-4820},
    Amedeo Romagnolo\inst{1,4}
    \orcid{0000-0001-9583-4339},
     Michela Mapelli\inst{1,4,5,6}
    \orcid{0000-0001-8799-2548} 
    }
    \authorrunning{L. Boco et al.}
    \institute{
    $^{1}$Universit\"at Heidelberg, Zentrum f\"ur Astronomie (ZAH), Institut f\"ur Theoretische Astrophysik, Albert Ueberle Str. 2, 69120, Heidelberg, Germany\\
    $^{2}$SISSA, Via Bonomea 365, I–34136 Trieste, Italy\\
    $^3$Department of Physics, University of Trento, Via Sommarive 14, 38123 Povo (TN), Italy\\
     $^4$Dipartimento di Fisica e Astronomia Galileo Galilei, Università di Padova, Vicolo dell’Osservatorio 3, I–35122 Padova, Italy\\
    $^5$Universit\"at Heidelberg, Interdiszipli\"ares Zentrum f\"ur Wissenschaftliches Rechnen, D-69120 Heidelberg, Germany\\
    $^6$INFN, Sezione di Padova, Via Marzolo 8, I--35131 Padova, Italy\\
    }

   \date{Received XXXX; accepted YYYY}

\abstract{After four observational runs, the Ligo-Virgo-Kagra  collaboration estimated a local binary black hole (BBH) merger rate density of $R_{0,\textrm{LVK}}\simeq 14-26\,\textrm{Gpc}^{-3}\,\textrm{yr}^{-1}$ within the 90\% credible interval. Some previous studies already pointed out that, when a realistic evolution of the metallicity-dependent cosmic star formation rate density (SFRD) is adopted, theoretical models predict a local BBH merger rate density that exceeds the observed value by at least a factor of $\sim 10$. We confirm and strengthen this claim by constructing an empirical model for the SFRD and metallicity evolution that includes a correction accounting for iron abundance. The adopted metallicity relation is flexible, enabling us to bracket the wide range of observational uncertainties. We show that even under the most conservative assumptions regarding the SFRD and the metallicity relation, the local BBH merger rate density is overestimated by a factor $> 10$. Attempts to reconcile the predicted and observed merger rates by modifying the metallicity-dependent SFRD alone would require unrealistically high metallicities ($Z>Z_\odot$) even in low-mass galaxies at high redshift. This finding indicates that revisions to the treatment of stellar and binary evolution are necessary to achieve consistency between theoretical predictions and observations. We suggest that even a modest steepening of the delay-time distribution might help alleviate this tension.}
  

   \maketitle

\defcitealias{Chruslinska2021}{Ch21}
\defcitealias{Chruslinska2025}{Ch25}
\defcitealias{Boco2021}{B21}

\section{Introduction}\label{sec:intro}
It has now been more than ten years since the first gravitational wave (GW) detection by the LIGO--Virgo--KAGRA (LVK)  collaboration. With the first four observing runs completed, we have started deriving with increasing precision the local binary black hole (BBH) merger rate density $R_{0,\textrm{LVK}}\simeq 14-26\,\textrm{Gpc}^{-3}\,\textrm{yr}^{-1}$, its evolution up to $z\gtrsim 1$, and the mass function of merging black holes \citep[BHs, ][]{Abac2026}. From a theoretical perspective, the local BBH merger rate density has been estimated in numerous studies by combining a so-called stellar and binary-evolution term -- that represents the outcomes of binary population synthesis codes -- with a metallicity-dependent star formation rate density SFRD$(z, Z)$, derived from either cosmological simulations \citep{Mapelli2017, Oshaughnessy2017, Lamberts2018, Mapelli2018, Artale2019, Levina2026} or empirical relations \citep{Belczynski2016, Lamberts2016, Cao2018, Elbert2018, Li2018, Boco2019, Chruslinska2019, Neijssel2019, Santoliquido2020, Boco2021, Santoliquido2022, Broekgaarden2022, Romagnolo2023,Romagnolo2025,Sgalletta2025}. 

The role of SFRD$(z, Z)$ is to provide a synthetic Universe within which binaries generated by population synthesis codes can evolve. It plays a fundamental role in determining the rate and properties of BBH mergers because the metallicity critically determines the fate of massive stellar binaries. Several studies \citep{Chruslinska2019, Boco2021, Chruslinska2021, Santoliquido2022, vanson2023, Sgalletta2025} that accurately modeled the $\textrm{SFRD}(z,Z)$ based on observations highlighted  that the metallicity distribution is not symmetric and features a pronounced tail towards low metallicities. This tail is primarily produced by low-mass galaxies and starbursts. The very existence of such a low-metallicity tail is relevant for computing BBH merger rates. The BBH merger efficiency, that is, the number of BBHs that merge within a Hubble time per unit of stellar mass formed, depends on the metallicity and reaches its maximum at low metallicities $Z\lesssim Z_\odot/3$ \citep{Giacobbo2018, Klencki2018,Chruslinska2019a, Iorio2023, vanson2025}. The metallicity distribution must therefore be well characterized for any robust computation of the BBH merger rate density. The shape of the metallicity distribution and the extension of the low-metallicity tail depend on the assumptions made in modeling galaxies, including the choice of the galaxy stellar mass function, the star-forming main sequence, the starburst fraction, and the metallicity relation \citep{Boco2021, Chruslinska2021}.

\cite{Sgalletta2025} used the binary population synthesis code \textsc{sevn} and an up-to-date metallicity relation \citep{Andrews2013} to show that the low-metallicity tail leads to an overestimation of the local BBH merger rates with respect to the observed rate $R_{0,\textrm{LVK}}$ by a factor $\gtrsim 10$, independently of the prescription adopted for the main sequence of star-forming galaxies. The authors concluded that modifications of the stellar and binary evolution term are required to reconcile the predicted and observed local merger rates. Possible solutions include suppressing specific BBH formation channels, such as those involving a common-envelope phase \citep[e.g.,][]{Olejak2021, Romagnolo2025}, or increasing the strength of natal kicks.

This prediction is even too optimistic: most calculations of the BBH merger rate density based on empirical metallicity scaling relations rely on oxygen abundances O/H, since oxygen is the primary tracer used to infer gas-phase metallicities in star-forming galaxies. However, line-driven stellar winds for $Z\gtrsim 0.1~Z_\odot$ are driven by iron lines, while at lower metallicities, their dependence on other driving elements is nearly completely unconstrained \citep[and references therein]{Romagnolo2026}. This implies that stellar and binary evolution are more sensitive to iron abundance than to oxygen. Oxygen-based metallicities are thus usually rescaled to a total metallicity $Z$ by using solar relative abundances as a benchmark. However, several observations of star-forming galaxies \citep{Steidel2014, Steidel2016, Topping2020, Sanders2021, Cullen2021, Strom2022} and of local quiescent galaxies \citep{Thomas2005, Thomas2010, Johansson2012, Alonso-Alvarez2025} show $\alpha$-enhancements up to $\textrm{O}/\textrm{Fe}\sim 4\,(\textrm{O}/\textrm{Fe})_\odot$. This enhancement arises because oxygen and iron have different enrichment timescales: oxygen is predominantly produced by core-collapse supernovae, while a substantial fraction of iron originates from Type Ia supernovae, which occur with a significant delay. Therefore, using solar abundance ratios to rescale oxygen and iron metallicities to a common global $Z$ might lead to incorrect results. Specifically, this rescaling results in an overestimation of the iron abundance in all the $\alpha$-enhanced galaxies, potentially affecting predictions of the BBH merger rates.

\cite{Chruslinska2024, Chruslinska2025} provided an empirical calibration of the relative oxygen-to-iron abundance ratio O/Fe as a function of the galaxy specific star formation rate (sSFR), which tracks the age of the stellar population. This relation should be incorporated into models that compute merger rates using oxygen-based metallicity relations. Implementing this correction would further reduce the inferred iron metallicity and would consequently increase the predicted BBH merger rates. This would exacerbate the discrepancy highlighted by \cite{Sgalletta2025}.

In this paper, we investigate the local BBH merger rate density $\mathcal{R}_0$ by constructing a synthetic Universe using the empirical model proposed by \cite{Boco2021}. We adopt the most conservative, sometimes even unrealistic, assumptions for the stellar mass function, the main sequence of star-forming galaxies, and the starburst distribution, with the aim of lowering $\mathcal{R}_0$ as much as possible. We assign a gas-phase metallicity to our mock galaxies using the fundamental metallicity relation (FMR) derived by \citet[hereafter \citetalias{Chruslinska2021}]{Chruslinska2021}, plus the correction of \citet[hereafter \citetalias{Chruslinska2025}]{Chruslinska2025}, thus accounting for iron abundances for the first time in an empirical model. The FMR of \citetalias{Chruslinska2021} provides a flexible parameterization that allows us to vary key properties, such as its overall normalization and its dependence on galaxy stellar mass and star formation rate. We vary the FMR parameters bracketing the range spanned by the main metallicity relations present in literature \citep{Lara-Lopez2010, Mannucci2010, Andrews2013, Salim2014, Torrey2018, Curti2020, Sanders2021, Curti2023, Nakajima2023}, and 
we investigate whether under reasonable metallicity variations, the local BBH merger rate density observed by LVK can be reproduced by models. We also investigate the impact of variations in the delay-time distribution between the formation of the stellar binary and the merger of the remnant BHs.

The paper is structured as follows: in Section \ref{sec:methods} we describe the methods we used to compute the BBH merger rate density, with particular focus on the conservative choices made to minimize $\mathcal{R}_0$; in Section \ref{sec:results} we present the results for reasonable variations in the FMR parameters; in Section \ref{sec:discussion} we reverse the approach and discuss the assumptions and prescriptions that should be used to reconcile the predicted and observed BBH merger rate; and in Section \ref{sec:conclusions} we summarize our main findings.

Throughout the paper we assume the Planck2018 cosmology, a flat Universe with $H_0\simeq0.68$ and $\Omega_m\simeq0.31$. We present results in terms of $[\textrm{X}/\textrm{H}]\equiv\log(\textrm{X}/\textrm{H})-\log(\textrm{X}/\textrm{H})_\odot$, where $\textrm{X}$ is a generic element, in the current context, either oxygen $\textrm{O}$ or iron $\textrm{Fe}$. The reference solar abundances are taken from \cite{Grevesse1998}, that is,  $12+\log(\textrm{O}/\textrm{H})_\odot=8.83$, $12+\log(\textrm{Fe}/\textrm{H})_\odot=7.5$, $\log(\textrm{O}/\textrm{Fe})_\odot=1.33$, and $Z_\odot=0.017$, for consistency with \citetalias{Chruslinska2025}.

\section{Methods}\label{sec:methods}
The BBH merger rate density at the cosmic time $t$ can be computed as
\begin{equation}
\mathcal{R}(t)=\int\,\textrm{d}\log Z\,\eta(Z)\int\textrm{d}t_d\,\frac{\textrm{d}p}{\textrm{d}t_d}(t_d|Z)\,\frac{\textrm{d}\dot M}{\textrm{d}V\,\textrm{d}\log Z}\left(t-t_d,Z\right)
\label{eq:merger_rate},
\end{equation}
where $Z$ is the metallicity, $t_d$ is the delay time between the formation of the binary and the merger of the BH remnants, $\textrm{d}p/\textrm{d}t_d$ is the delay-time distribution, possibly dependent on metallicity, $\eta(Z)$ is the BBH merger efficiency, defined as the number of BBHs formed per unit of star formation whose delay time is shorter than a Hubble time, and $\textrm{d}\dot M/\textrm{d}V\,\textrm{d}\log Z$ is the metallicity-dependent star formation rate density $\textrm{SFRD}(z,Z)$, computed at the time $t-t_d$. The merger efficiency and delay-time distribution are usually computed from the outcomes of binary population synthesis codes. In this work, we mainly adopt the results of the binary population synthesis code \textsc{sevn}, the fiducial model by \cite{Iorio2023}, with a common-envelope efficiency $\alpha_{\textrm{CE}}=1$. However, in Section \ref{sec:results_alpha}, we show the effect of varying the parameter $\alpha_{\textrm{CE}}$, and in Section \ref{sec:discussion_comparison}, we compare our results with other studies using different binary population synthesis codes. We assume a binary fraction $f_{\textrm{bin}}=0.5$, which is a conservative assumption as massive stars tend to feature higher binary fractions $>0.5$ \citep{Sana2012}.

The metallicity-dependent SFRD is computed by integrating the star formation rate (SFR) of all the galaxies in a given cosmological volume \citep[][]{Boco2021},
\begin{equation}
\begin{split}
\frac{\textrm{d}\dot M}{\textrm{d}V\,\textrm{d}\log Z}(t,Z)&=\int_{\log M_{\star, \textrm{min}}}\textrm{d}\log M_\star\,\frac{\textrm{d}N(M_\star,t)}{\textrm{d}V\,\textrm{d}\log M_\star}\int\textrm{d}\log\psi\\
&\times\,\psi\, \frac{\textrm{d}p}{\textrm{d}\log\psi}(\log\psi|t,M_\star)\,\frac{\textrm{d}p}{\textrm{d}\log Z}(\log Z|M_\star,\psi)
\end{split}
\label{eq:SFRD},
\end{equation}
where $\textrm{d}N/\textrm{d}V\,\textrm{d}\log M_\star$ is the star-forming galaxy stellar mass function (GSMF), $\log M_{\star, \textrm{min}}$ is the minimum stellar mass accounted for in the integration, $\psi$ is the SFR, $\textrm{d}p/\textrm{d}\log\psi$ is the distribution of SFR at a given mass and cosmic time and stellar mass, and $\textrm{d}p/\textrm{d}\log Z$ is the metallicity distribution as a function of the galaxy stellar mass and SFR. We describe these elements in detail in the sections below.

\subsection{Galaxy stellar mass function (GSMF)}
The GSMF $\textrm{d}N/\textrm{d}V\,\textrm{d}\log M_\star$ represents the number density of galaxies per bin of stellar mass at different redshifts. It has been determined by many observational studies \citep{Ilbert2013, Muzzin2013, Tomczak2014, Davidzon2017, Weaver2023, Shuntov2025}. We adopt the recent determination of the GSMF of star-forming galaxies by \cite{Weaver2023}, derived from the COSMOS2020 catalog. To compute the BBH merger rate, the faint end of the GSMF may be of some importance because low-mass galaxies are poorer in metals and might host a significant number of BBH mergers. Therefore, the choice of the integration limit $\log M_{\star, \textrm{min}}$ in equation \eqref{eq:SFRD} is relevant. Since we want to provide a conservative estimate of $\mathcal{R}_0$, we choose $M_{\star, \textrm{min}}=10^8\,\textrm{M}_\odot$, meaning that the contribution to the BBH merger rate density from all the galaxies with mass $M_\star<10^8\,\textrm{M}_\odot$ is not taken into account.

\subsection{SFR distribution}
The galaxy SFR distribution $\textrm{d}p/\textrm{d}\log\psi$ at a given stellar mass and cosmic time can be computed from the main sequence of star-forming galaxies, a well-know correlation between stellar mass and SFR, which has been determined observationally and theoretically in several works \citep{Daddi2007, Rodighiero2011, Rodighiero2015,
Speagle2014, Whitaker2014, Schreiber2015, Mancuso2016b, Dunlop2017, Bisigello2018, Pantoni2019, Lapi2020, Popesso2023, Bosi2025}. We adopt the relation by \cite{Popesso2023}, which is one of the most  recent observational determinations in a wide range of masses and redshift.

Although the main sequence is the locus in the $\log\psi-\log M_\star$ plane where most star-forming galaxies are concentrated, the SFR distribution at a fixed stellar mass and cosmic time is more complex. Many authors pointed out that this distribution tends to be bimodal, with a prominent peak representing main-sequence galaxies, and a secondary peak at higher SFR representing starburst galaxies \citep{Rodighiero2011, Bethermin2012, Sargent2012, Ilbert2015, Schreiber2015, Caputi2017, Bisigello2018, Rinaldi2022, Rinaldi2025}. This distribution is usually modeled with a double Gaussian in $\log\psi$, with the relative amplitude of the two Gaussians controlling the relative abundance of the two populations \citep{Sargent2012, Boco2021}. Several authors \citep{Rodighiero2011, Bethermin2012, Sargent2012, Ilbert2015, Schreiber2015} estimated the fraction of starburst galaxies $f_{\textrm{SB}}$ to be $\sim 3\%$. However, most of these estimates were limited to the most massive galaxies $M_\star\gtrsim 10^{10}\,\textrm{M}_\odot$ at redshift $z<3$. Recent studies \citep{Caputi2017, Bisigello2018, Rinaldi2022, Rinaldi2025} extended the mass and redshift range and pointed out that the starburst fraction tends to increase steeply at lower masses and higher redshift, reaching values of $f_{\textrm{SB}}\gtrsim 90\%$ for $M_\star\lesssim 10^{7}\,\textrm{M}_\odot$ \citep{Rinaldi2025}. However, even galaxies with $M_\star\gtrsim 10^8\,\textrm{M}_\odot$ feature $f_{\textrm{SB}}\sim 20-40\%$ at $z\sim 3-6$. Since the starburst SFR is higher by $\sim 3-10$ than that of main-sequence galaxies and they are typically poorer in metals, these high values of $f_{\textrm{SB}}$ might substantially increase the BBH merger rates, as shown by \citetalias{Chruslinska2021}. Here, as we aim to produce a conservative estimate of $\mathcal{R}_0$, we completely ignore the starburst population. We set $f_{\textrm{SB}}=0\%$.

\subsection{Metallicity distribution}\label{sec:metallicity}

\subsubsection{The fundamental metallicity relation}\label{sec:FMR}
The metal content of the cold star-forming gas in galaxies is of great importance for assessing the masses and merger rates of BBHs because the strength of line-driven winds for massive stars critically depends on metallicity\footnote{Cool supergiant winds are instead not driven by lines, but likely by factors such as near-surface convective boil-off \citep{Fuller2024} and turbulence \citep{Kee2021}. Their strength therefore remains similar in different metallicity environments \citep{Antoniadis2025}.}. There are several observational relations between galaxy properties and gas-phase metallicity. These are usually given in terms of the oxygen abundance $12+\log(\textrm{O}/\textrm{H})$, whose emission lines are among the most luminous in the spectra of HII regions. Throughout this work, we define $Z_{\textrm{O}/\textrm{H}}\equiv12+\log(\textrm{O}/\textrm{H})$ and we assign a metallicity\footnote{When talking about the FMR, we use the term metallicity, or gas-phase metallicity, but we refer to the oxygen abundance $Z_{\textrm{O}/\textrm{H}}$.} to galaxies based on the fundamental metallicity relation, a relation between gas-phase metallicity, galaxy stellar mass, and SFR \citep{Lara-Lopez2010, Mannucci2010, Mannucci2011, Andrews2013, Zahid2014b, Hunt2016, Cresci2019, Curti2020, Sanders2021, Boco2021, Chruslinska2021, Curti2023, Nakajima2023}. The FMR has no explicit redshift dependence: the redshift evolution of the metallicity at fixed stellar mass is implicitly accounted for through the metallicity-SFR anticorrelation. At higher redshift, the average SFR of galaxies increases, and thus, the metallicity derived from the FMR decreases. This feature allows to use the FMR to compute the metallicity even at high z, without the need of extrapolating the relation, as shown by \cite{Boco2021}. The assumption of redshift independence of the FMR has been tested by multiple authors \citep{Mannucci2010, Hunt2016, Curti2023, Nakajima2023}, and its validity is somewhat dependent on the chosen parameterization. In particular, through a comparison with JADES and CEERS data, the FMR derived by \cite{Curti2020} has recently been proved to hold up to $z\sim 4-6$, while that by \cite{Andrews2013} holds up to $z\sim 8$ \citep{Curti2023, Nakajima2023}.

In this work, we use the FMR parameterization introduced by \citetalias{Chruslinska2021}. This is a flexible parameterization that makes use of the functional form of \cite{Curti2020}, but with some free parameters that allow us to test the dependence of $\mathcal{R}_0$ on different metallicity relations. The functional form of the FMR is the following:
\begin{equation}
Z_{\textrm{O}/\textrm{H}}=Z_{\textrm{O}/\textrm{H}, 0}-\frac{\gamma}{\beta}\,\log\left(1+\left(\frac{M_\star}{M_{0,SFR}}\right)^{-\beta}\right),
\end{equation}
where $Z_{\textrm{O}/\textrm{H}, 0}$ is a normalization, and the dependence on the SFR is enclosed in $\log M_{0,SFR}\equiv m_0+m_1\,\log\psi$. In the low-mass regime ($M_\star<<M_{0,SFR}$), it reduces to a linear relation in log,
\begin{equation}
Z_{\textrm{O}/\textrm{H}}\propto\gamma\,(\log M_\star-m_1\log\psi)\propto\gamma\,\log M_\star-\nabla_{\textrm{FMR},0}\log\psi,
\end{equation} 
where $\gamma$ is the slope of the correlation between metallicity and stellar mass, while $\nabla_{\textrm{FMR},0}\equiv\gamma\,m_1$ is the slope of the anticorrelation between metallicity and SFR. This is the same functional form
derived by \cite{Mannucci2010, Mannucci2011, Andrews2013, Hunt2016}. \citetalias{Chruslinska2021} demonstrated that in the low-mass regime, $\gamma$ and $\nabla_{\textrm{FMR},0}$ are connected to the slope of the mass-metallicity relation (MZR), $a_{\textrm{MZR}}$, and the slope of the main sequence $a_{\textrm{MS}}$, computed at $z=0$, by the following relation: $\gamma=\nabla_{\textrm{FMR},0}\,a_{\textrm{MS}} + a_{\textrm{MZR}}$. Since we selected the main-sequence determination of \cite{Popesso2023}, which has a faint-end slope $a_{\textrm{MS}}\sim 1$ at $z=0$, this relation reduces to $\gamma=\nabla_{\textrm{FMR},0} + a_{\textrm{MZR}}$. The parameter $m_0$ can also be derived from the shape of the MZR and main sequence, but with a more complicated expression (see \citetalias{Chruslinska2021} for more details). Thus, the shape of the FMR can be entirely derived from the local MZR and main sequence when a value for the parameter $\nabla_{\textrm{FMR},0}$ is specified. We vary the parameters $a_{\textrm{MZR}}$ and $\nabla_{\textrm{FMR},0}$ in reasonable ranges and use them to derive the FMR parameters $\gamma$, $m_1\equiv\nabla_{\textrm{FMR},0}/\gamma$ and $m_0$. We also vary the overall normalization $Z_{\textrm{O}/\textrm{H},0}$. The choices for these parameters are summarized in Table \ref{table:parameters}. In Appendix \ref{sec:appendix} we show that these choices bracket most of the values found in the literature. Since the FMR is redshift independent, we use it to compute metallicities at high redshift.

\begin{table}
\centering
\footnotesize 
\caption{Description of the parameters}
\begin{tabular}{p{0.12\columnwidth}  p{0.5\columnwidth}  p{0.3\columnwidth}|}
\hline
Parameter & Description & Values \\
\hline
$a_{\textrm{MZR}}$ & Slope of the $z=0$ MZR & 0.15--0.3--0.6 \\

$\nabla_{\textrm{FMR},0}$ &
Strength of $Z$--SFR anti-correlation &
0.2--0.3 \\

$Z_{\textrm{O}/\textrm{H},0}$ &
FMR normalization &
8.8--9.0--9.2 \\

$\gamma$ &
Strength of $Z-M_\star$ correlation &
$a_{\textrm{MZR}}+a_{\textrm{MS}}$ \\

$m_0$ &
Position of the FMR knee &
Eq.~6 in \citetalias{Chruslinska2021} \\

$m_1$ &
Dependence of knee on SFR &
$\nabla_{\textrm{FMR},0}/\gamma$ \\

$\beta$ &
Width of the FMR knee &
2.1 \citep{Curti2020} \\
\hline
\end{tabular}
\parbox{0.95\columnwidth}{
Notes: Description of the parameters and values. We vary $a_\textrm{MZR}$, $\nabla_{\textrm{FMR},0}$, and $Z_{\textrm{O}/\textrm{H},0}$ among the values reported in the Table. $\gamma$, $m_0$, $m_1$, and $\beta$ are computed as written in the table.
}
\label{table:parameters}
\end{table}

Finally, the scatter used around the FMR, $\sigma_\textrm{FMR}$, is important for calculating the merger rate because it affects the extension of the low-metallicity tail. A higher value of $\sigma_\textrm{FMR}$ implies more star formation at low metallicity, and thus, a larger $\mathcal{R}_0$. Various FMR estimations reported values of $\sigma_\textrm{FMR}$ in the range $0.05-0.15$. Despite recent JADES and CEERS data \citep{Nakajima2023, Curti2023} confirmed the validity of the FMR up to $z\sim4-8$, the scatter around this relation seems to increase at high redshift. We use $\sigma_\textrm{FMR}=0.05$ at all redshifts to minimize the value of $\mathcal{R}_0$. Many models that compute merger rates describe the metallicity in a cosmic-averaged way by selecting an average metallicity at each redshift $\langle Z(z)\rangle$ and introducing a scatter around it, either with a log-normal distribution or with a skewed log-normal. We caution that $\sigma_\textrm{FMR}$ reported here is not the same $\sigma$ as reported in those models. $\sigma$ of the cosmic-averaged models accounts for the overall metallicity dispersion at a fixed redshift, while $\sigma_\textrm{FMR}$ is meant to render the possible dispersion of the metallicity around the FMR, that is, the dispersion for galaxies with the same stellar mass and SFR. The overall dispersion at a given redshift, $\sigma$, is way larger than $\sigma_\textrm{FMR}$, and it is driven by the metallicities of individual galaxies with different $M_\star$ and SFR (see the discussion in \ref{sec:discussion_comparison}).

\subsubsection{Iron and oxygen abundances}\label{sec:iron_rescaling}

Most of the studies computing merger rates use the total metal mass fraction $Z$ in equation \eqref{eq:merger_rate}. However, as mentioned in Section \ref{sec:intro}, this might be misleading. The abundances of iron-group elements regulate the stellar opacity and the strength of line-driven winds. Therefore, the results of binary population synthesis codes, such as the merger efficiency $\eta(Z)$ and delay-time distribution $\textrm{d}p/\textrm{d}t_d(t_d|Z)$, are actually most affected by the iron abundance rather than by the total metallicity. On the other hand, as highlighted above, observations of star-forming galaxies typically constrain the oxygen abundance. The iron and oxygen abundances are generally converted into total metallicity using the solar relative abundances as a reference scale (e.g., $Z=Z_\odot\,10^{[\textrm{O}/\textrm{H}]}$). 

\begin{figure}
    \centering
    \includegraphics[width=0.5\textwidth]{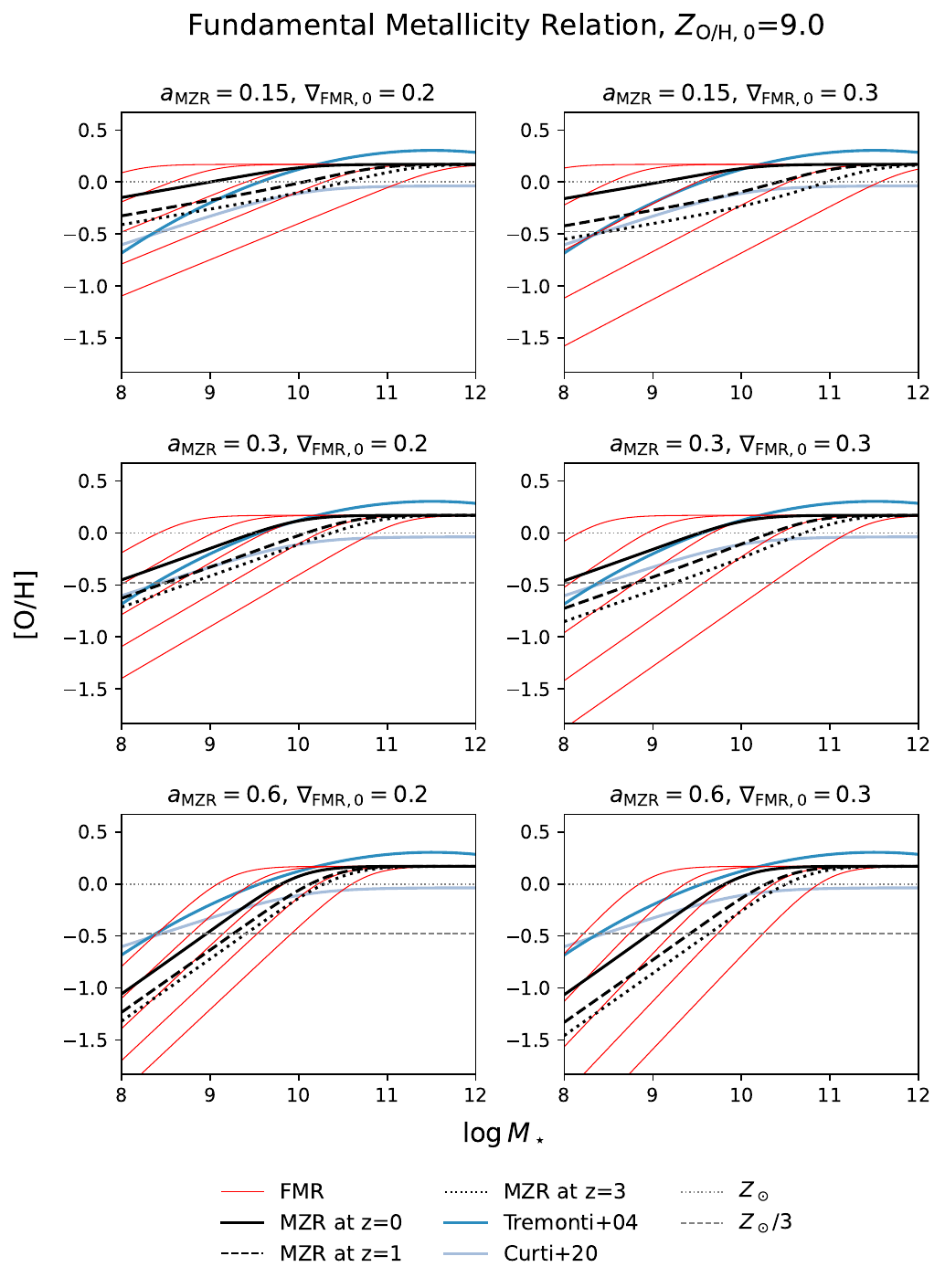}
    \caption{FMR and derived MZR for $Z_{\textrm{O}/\textrm{H},0}=9.0$, in terms of $[\textrm{O}/\textrm{H]}$. The different panels show different values of $a_{\textrm{MZR}}$ and $\nabla_{\textrm{FMR},0}$. The thin red lines show the FMR at different SFR, $\log\psi=-3,\,-1.5,\,0,\,1.5,$ and $3$ from the top to the bottom line. The black lines (solid, dashed and dotted) show the derived MZR at $z=0,\,1,\,$ and $3$, respectively. The blue lines are $z=0$ MZR from the literature: \cite{Tremonti2004} in dark blue, and \cite{Curti2020} in steel blue. The dashed and dotted horizontal gray lines represent $Z_\odot/3$ and $Z_\odot$.}
    \label{fig:metallicity_relation}
\end{figure}

However, oxygen and iron have different enrichment timescales, with oxygen being produced mostly by massive stars, and thus tracing the SFR, while iron receiving a substantial contribution by type Ia supernovae (SNIa), which might have a longer delay time. Some galaxies, especially the youngest, are therefore $\alpha$-enhanced with respect to the solar abundance. High values of the $\alpha$-enhancement have been measured in high-redshift star-forming galaxies \citep{Steidel2014, Steidel2016, Topping2020, Sanders2021, Cullen2021, Strom2022} and in the stellar metallicity of local ellipticals \citep{Thomas2005, Thomas2010, Johansson2012, Alonso-Alvarez2025}, which are formed in short and intense bursts of star formation and are quenched before SNIa had the time to enrich the interstellar medium with iron \citep{Chiappini1997, Gallazzi2006, Thomas2005, Thomas2010, Johansson2012, Courteau2014, Pezzulli2016, Bellstedt2024, Alonso-Alvarez2025}. 

\cite{Chruslinska2024} and \citetalias{Chruslinska2025} derived a handy empirical relation between the O/Fe ratio and the galaxy specific SFR ($\textrm{sSFR}\equiv\psi/M_\star$), as the latter quantity is a proxy for the galaxy age. We use the O/Fe-sSFR relation by \citetalias{Chruslinska2025}, in the 'Mixed' Fe enrichment case. We use this relation to evaluate $[\textrm{O}/\textrm{Fe}]$ as a function of the stellar mass and SFR of the galaxy, and we compute the iron abundance as $[\textrm{Fe}/\textrm{H}]=[\textrm{O}/\textrm{H}]-[\textrm{O}/\textrm{Fe}]$, where $[\textrm{O}/\textrm{H}]=Z_{\textrm{O}/\textrm{H}}-Z_{\textrm{O}/\textrm{H},\odot}$ is derived from the FMR. The total metallicity $Z$ in equations \eqref{eq:merger_rate} and \eqref{eq:SFRD} is then computed by rescaling the iron abundance as $Z=Z_\odot\,10^{[\textrm{Fe}/\textrm{H}]}$. The metallicity distribution $\textrm{d}p/\textrm{d}\log Z$ is a Gaussian distribution, with a scatter $\sigma_{\textrm{FMR}}\sim 0.05$, as discussed in Section \ref{sec:metallicity}. The rescaling to iron abundance through the relation by \citetalias{Chruslinska2025} might introduce another source of scatter. We ignore this possible additional scatter as it would contribute to the low-metallicity tail of the $\textrm{SFRD}(z,Z)$ and would thus increase $\mathcal{R}_0$. 

\begin{figure}
    \centering
    \includegraphics[width=1.0\linewidth]{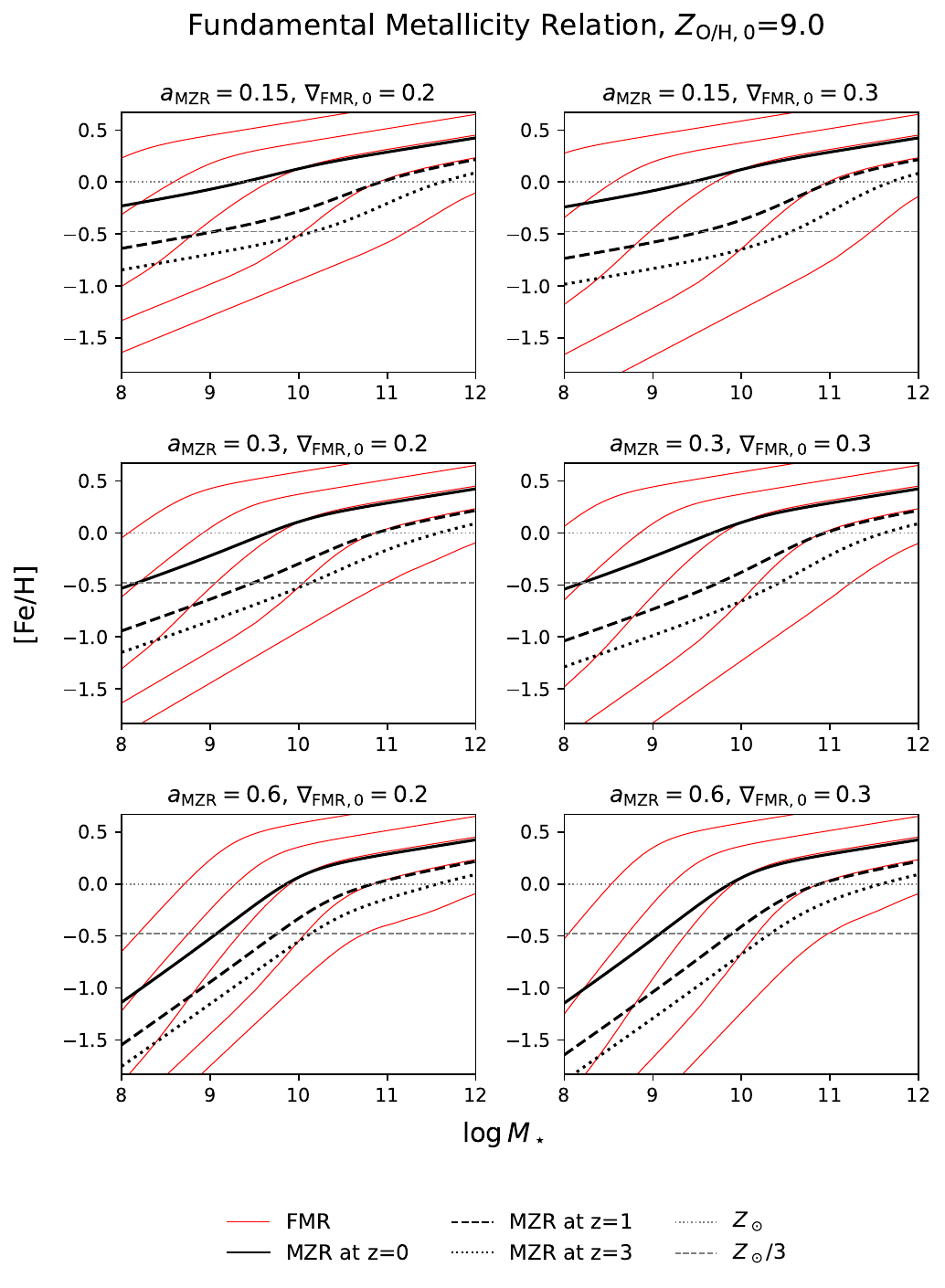}
    \caption{Same as Figure \ref{fig:metallicity_relation}, but for $[\textrm{Fe}/\textrm{H]}$.}
    \label{fig:metallicity_relation_Fe}
\end{figure}

\section{Results}\label{sec:results}
\subsection{The fundamental metallicity relation}
Figure \ref{fig:metallicity_relation} shows the FMR in terms of $[\textrm{O}/\textrm{H]}$, derived for $a_{\textrm{MZR}}=0.15,\,0.3\,,0.6$ and $\nabla_{\textrm{FMR},0}=0.2,\,0.3$. The normalization $Z_{\textrm{O}/\textrm{H},0}$ is fixed at $9.0$, representing the middle of the three values considered, $8.8,\,9.0,\,9.2$. The other two values would correspond to a shift of $\pm 0.2$ dex of the whole relation. The slope of the correlation with the stellar mass, $\gamma$, and the separation between the curves at different SFR, $m_1$, are regulated by the two input parameters $a_{\textrm{MZR}}$ and $\nabla_{\textrm{FMR},0}$. The figure also shows the MZR associated with that FMR and its average redshift evolution,
\begin{equation}
Z_{\textrm{MZR}}(M_\star,z)=\int\textrm{d}\log\psi\,\frac{\textrm{d}p}{\textrm{d}\log\psi}(\log\psi|z,M_\star)\,Z_{\textrm{FMR}}(M_\star,\psi).
\end{equation}
For comparison, we also show the $z=0$ MZR by \cite{Tremonti2004, Andrews2013} and \cite{Curti2020}. The redshift evolution of the MZR is regulated by the parameter $m_1$, which sets the separation between the curves at different SFR, and thus, it is governed by the input $\nabla_{\textrm{FMR},0}$. Figure \ref{fig:metallicity_relation_Fe} displays the same relations, but in terms of $[\textrm{Fe}/\textrm{H}]$, that is, with the \citetalias{Chruslinska2025} $\textrm{O}/\textrm{Fe}$ correction.

\begin{figure}
    \centering
    \includegraphics[width=1.0\linewidth]{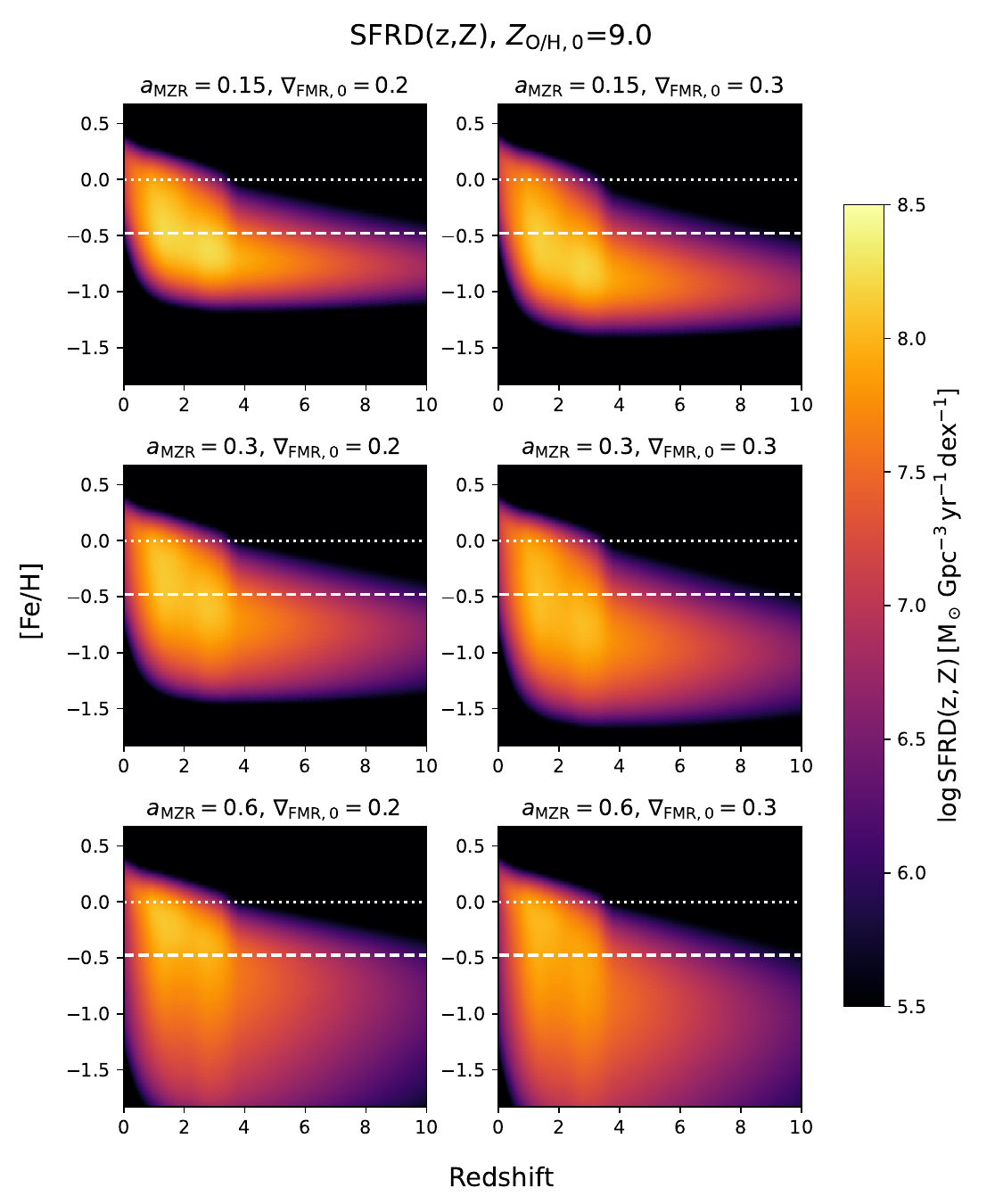}
    \caption{Cosmic SFRD per metallicity bin ($\textrm{d}\dot M/\textrm{d}V\,\textrm{d}\log Z$) for $Z_{\textrm{O}/\textrm{H},0}=9.0$ as a function of $z$ and $[\textrm{Fe}/\textrm{H]}$. The different panels show different values of $a_{\textrm{MZR}}$ and $\nabla_{\textrm{FMR},0}$. The dashed and dotted horizontal whit lines represent $Z_\odot/3$ and $Z_\odot$.}
    \label{fig:SFRD}
\end{figure}

The $\textrm{SFRD}(z,Z)$ associated with the different metallicity relations is shown in Figure \ref{fig:SFRD} in terms of $[\textrm{Fe}/\textrm{H}]$. It features a quite broad peak at $1<z<3$, the cosmic noon. The average $[\textrm{Fe}/\textrm{H}]$ evolves with redshift: at $z=0$ it ranges from $\sim Z_\odot/2$ to slightly supersolar, depending on the metallicity parameters chosen, at $z\sim4$ it ranges from $\sim Z_\odot/10$ to $\sim Z_\odot/5$. Models with different $Z_{\textrm{O}/\textrm{H},0}$ are simply shifted upward or downward by $0.2$ dex. This redshift evolution is partly due to the decrease in the average stellar mass and the increase in the galaxy SFR at higher redshift, and partly to the $[\textrm{Fe}/\textrm{O}]$ correction, which is larger for high-z galaxies, featuring a higher average sSFR. At $z>4$, the redshift evolution is shallower because the increase in the SFR with redshift saturates. The distribution is not symmetric, but features a low-metallicity tail that becomes more extended as $a_{\textrm{MZR}}$ and $\nabla_{\textrm{FMR},0}$ increase. A larger $a_{\textrm{MZR}}$ means that low-mass galaxies are more metal poor, resulting in a more extended low-metallicity tail. A larger $\nabla_{\textrm{FMR},0}$ means a stronger anticorrelation with SFR. This implies a more prominent low-metallicity tail, originated by more star-forming galaxies, and a slightly faster redshift evolution of the metallicity.

\subsection{The BBH merger rate density}

\begin{figure}
    \centering
    \includegraphics[width=1.0\linewidth]{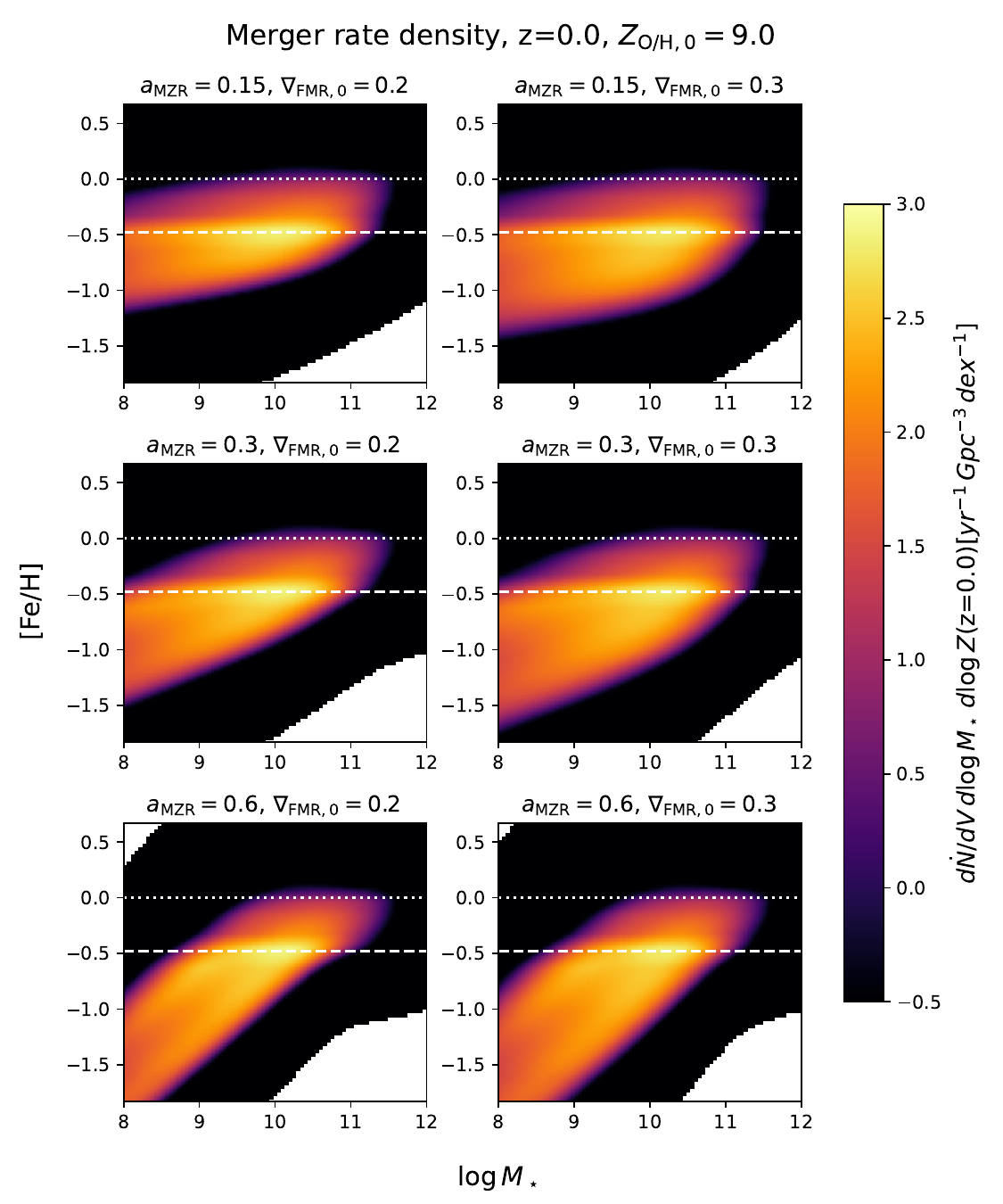}
    \caption{BBH merger rate density per metallicity and stellar mass bin ($\textrm{d}\dot N/\textrm{d}V\,\textrm{d}\log Z\,\textrm{d}\log M_\star$) for $Z_{\textrm{O}/\textrm{H},0}=9.0$, computed at $z=0$, as a function of $\log M_\star$ and $[\textrm{Fe}/\textrm{H]}$. The different panels show different values of $a_{\textrm{MZR}}$ and $\nabla_{\textrm{FMR},0}$. The dashed and dotted horizontal white lines represent $Z_\odot/3$ and $Z_\odot$. The contribution to the BBH merger rate density comes almost exclusively from $Z<Z_\odot$ and is larger for $Z<Z_\odot/3$, due to the behavior of the merger efficiency $\eta(Z)$. The $z=0$ BBH merger rate density is reduced as $a_{\textrm{MZR}}$ and $\nabla_{\textrm{FMR},0}$ decrease.}
    \label{fig:WSFRD_efficiency}
\end{figure}

The local BBH merger rate density $\mathcal{R}_0$ receives contributions not only from the local SFRD, but also from BBHs formed at higher redshift that merge at $z=0$. In particular, the high SFRD at cosmic noon contributes significantly to the number of mergers occurring at $z=0$. The convolution of the SFRD with the delay-time distribution multiplied by the efficiency $\eta(Z)$ gives the merger rate density of BBH. 

\begin{figure}
    \centering
    \includegraphics[width=0.85\linewidth]{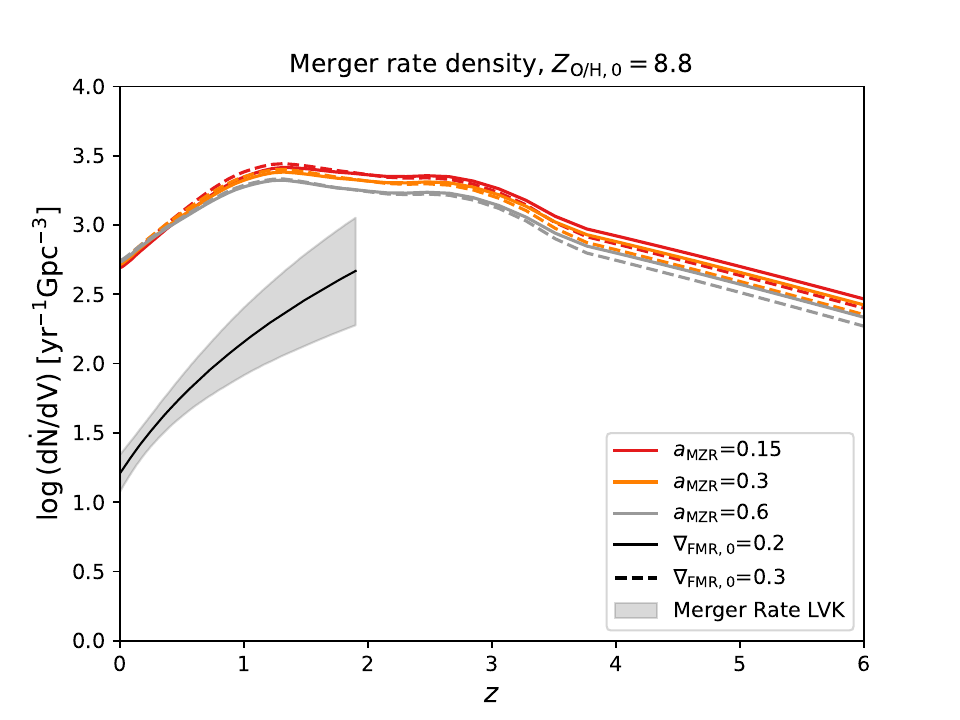}
    \includegraphics[width=0.85\linewidth]{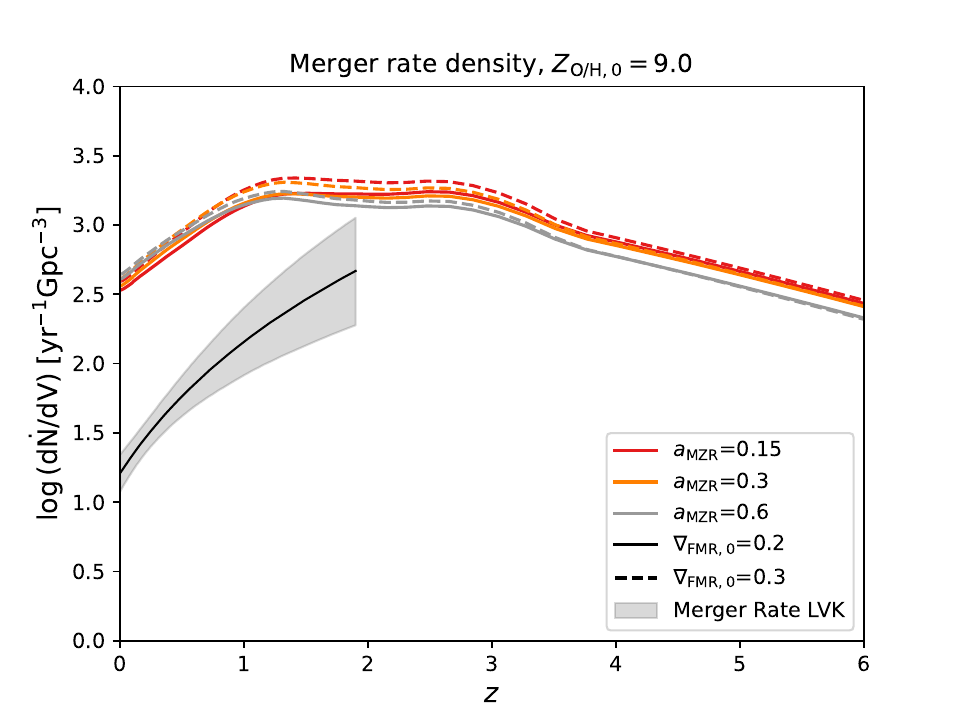}
    \includegraphics[width=0.85\linewidth]{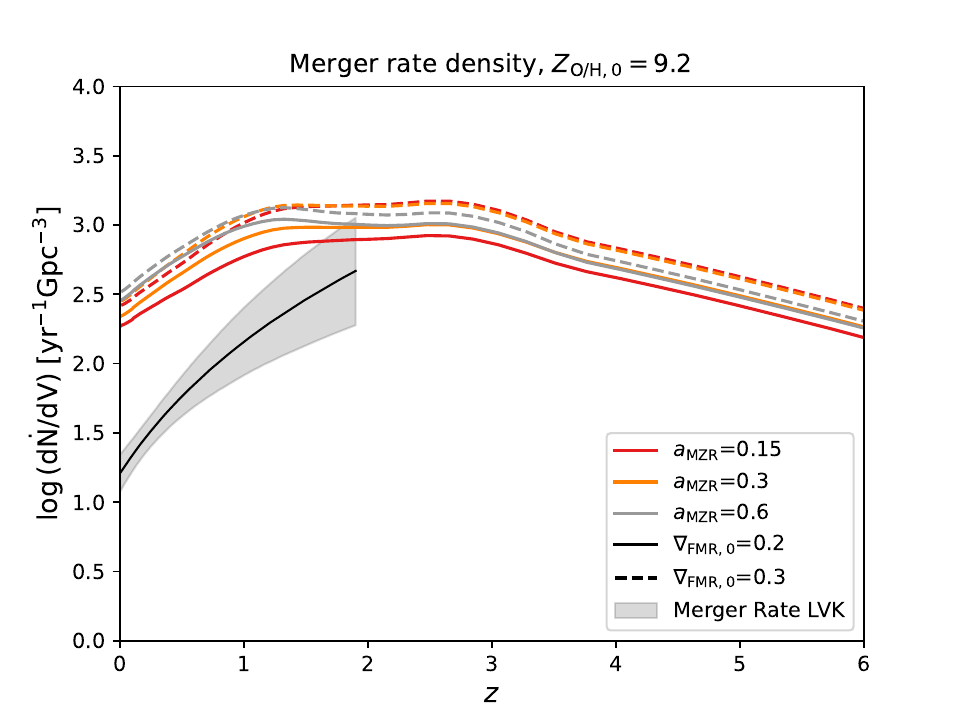}
    \caption{ Merger rate density as a function of redshift for $Z_{\textrm{O}/\textrm{H},0}=8.8,\,9.0,$ and $9.2$ (top, middle, and bottom panels), $a_{\textrm{MZR}}=0.15,\,0.3, and\,0.6$ (red, orange, and gray lines), and $\nabla_{\textrm{FMR},0}=0.2$ and $0.3$ (solid and dashed lines). The black shaded area shows the merger rate density measured by LVK.}
    \label{fig:merger_rates}
\end{figure}

Figure \ref{fig:WSFRD_efficiency} shows the merger rate density at $z=0$ per unit of galaxy stellar mass and metallicity $\textrm{d}\dot N/\textrm{d}V\,\textrm{d}\log M_\star\,\textrm{d}\log Z$ and indicates the stellar mass and metallicity of the galaxies where BBHs that merge at $z=0$ are formed. The integral of this quantity over $M_\star$ and $Z$ gives the local BBH merger rate density $\mathcal{R}_0$. The dependence of the merger rate density on metallicity reflects the behavior of the efficiency factor $\eta(Z)$. This factor is derived from binary population synthesis codes and usually features an almost constant value $\sim10^{-4.5}$ at $Z<Z_\odot/3$, while it rapidly decreases for $Z\gtrsim Z_\odot/3$. Thus, the local BBH merger rate density is mostly determined by the amount of star formation that occurs at metallicities lower than this threshold. The key to reducing the BBH merger rate for our default binary evolution models is to minimize the amount of star formation occurring at $Z<Z_\odot/3$ . Variations with lower $a_{\textrm{MZR}}$, lower $\nabla_{\textrm{FMR},0}$ and larger $Z_{\textrm{O}/\textrm{H},0}$ feature a less extended tail at low metallicity and imply lower $\mathcal{R}_0$ values.

Figure \ref{fig:merger_rates} shows the BBH merger rate density as a function of redshift for all the model variations. This figure shows that, independently of the chosen metallicity parameters, the local BBH merger rate density estimated by the models is well above the 90\% credible interval inferred by the LVK collaboration. We note that the slope is also quite flatter than that predicted by observations. 

\begin{figure*}
    \centering
    \includegraphics[width=0.9\textwidth]{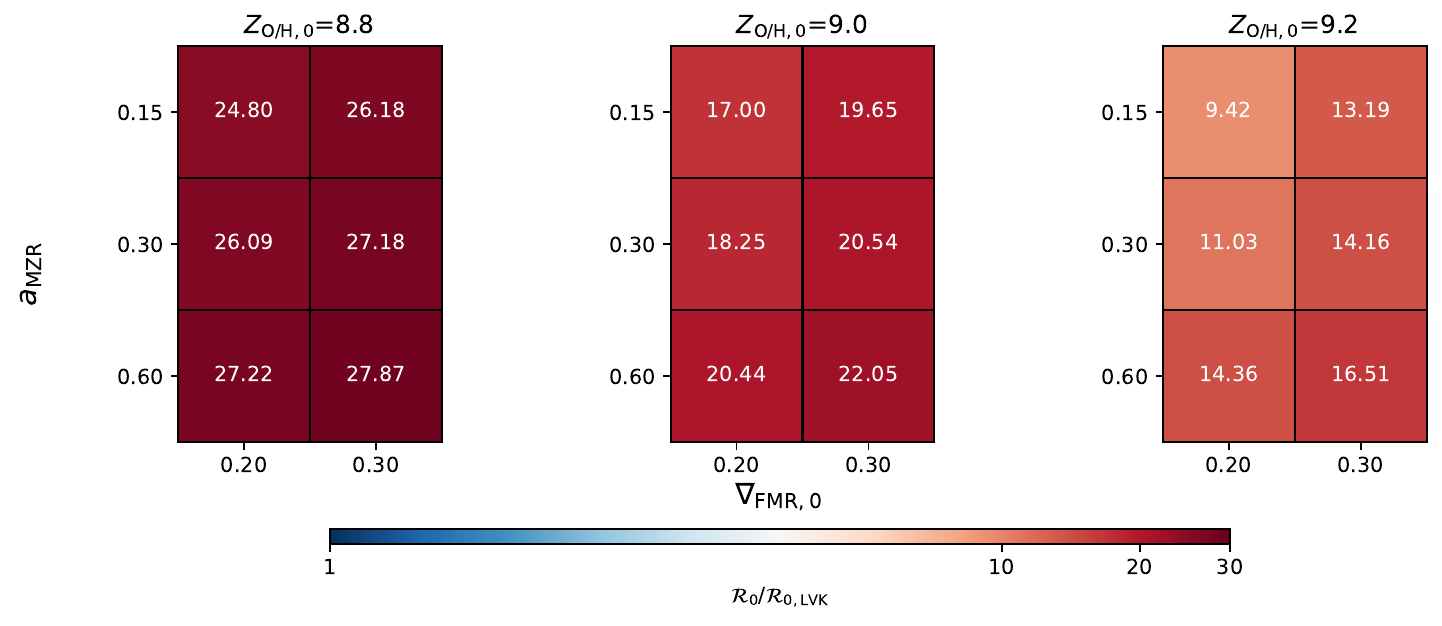}
    \caption{Matrix showing the ratio $\mathcal{R}_0/\mathcal{R}_{0,\textrm{LVK}}$ for $Z_{\textrm{O}/\textrm{H},0}=8.8,\,9.0,$ and $9.2$ (left, middle, and right tables), $a_{\textrm{MZR}}=0.15,\,0.3,$ and $0.6$ (y-axis), and $\nabla_{\textrm{FMR},0}=0.2$ and $0.3$ (x-axis). The color scale is set so that it is blue for $\mathcal{R}_0/\mathcal{R}_{0,\textrm{LVK}}=1$ and remains blue-shaded for $\mathcal{R}_0/\mathcal{R}_{0,\textrm{LVK}}\leq 2$, which corresponds to a distance of $\sim 6\sigma$ from the interval provided by LVK. For $\mathcal{R}_0/\mathcal{R}_{0,\textrm{LVK}}> 2$ the color becomes increasingly redder. No reasonable value of the FMR parameters can match $\mathcal{R}_{0,\textrm{LVK}}$.}
     \label{fig:table}
\end{figure*}

Figure \ref{fig:table} displays a matrix indicating the ratio of the theoretical estimation and the observed local rate as a function of the parameters of the metallicity relation. Even in the most optimistic case, for $Z_{\textrm{O}/\textrm{H},0}=9.2$, $a_{\textrm{MZR}}=0.15$, $\nabla_{\textrm{FMR},0}=0.2$, this ratio is $\sim 10$.

\begin{figure*}
     \centering
     \includegraphics[width=0.9\textwidth]{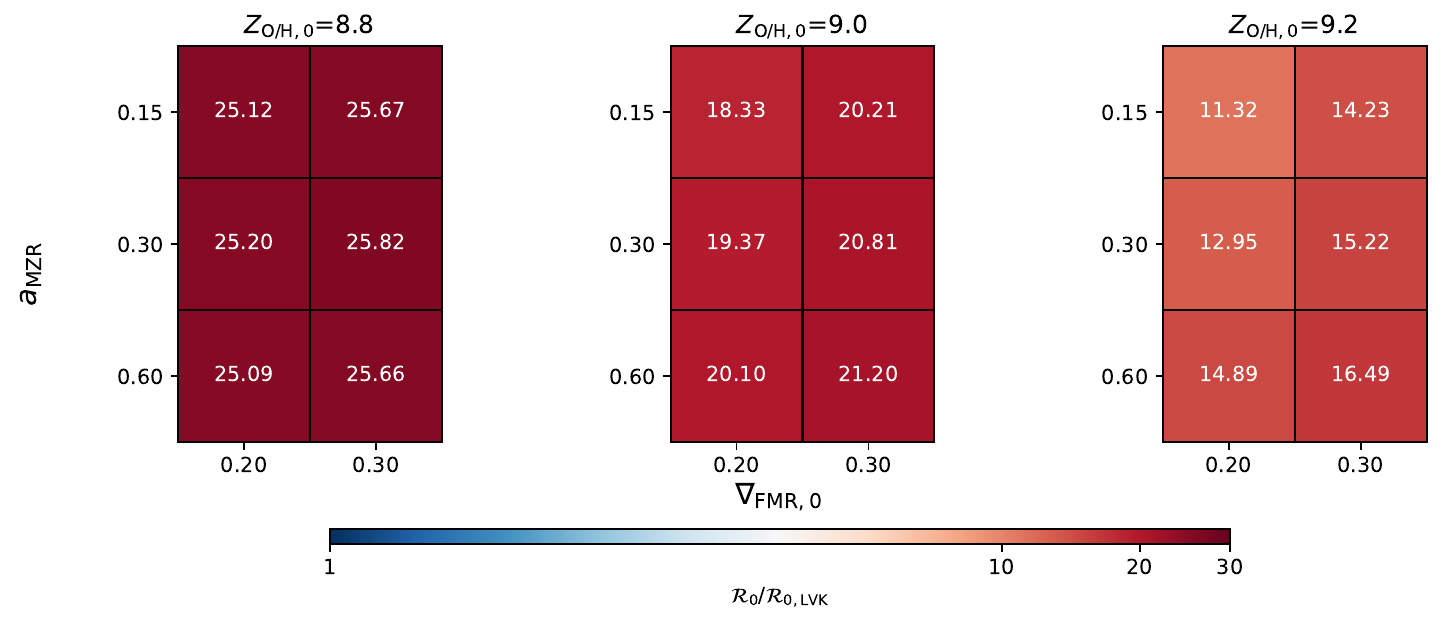}
     \vspace{0.2cm} 
     \includegraphics[width=0.9\textwidth]{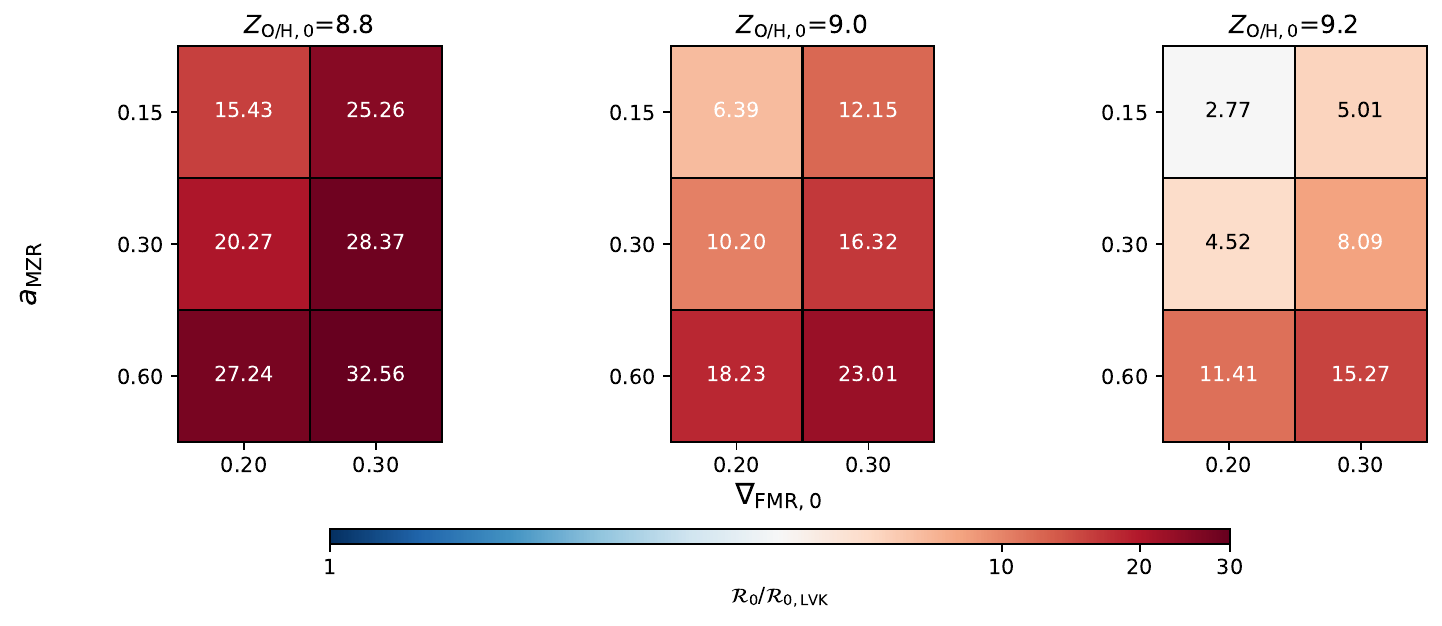}
   \caption{Same as Figure \ref{fig:table}, but for $\alpha_{\textrm{CE}}=0.1$ (top panel) and $\alpha_{\textrm{CE}}=10$ (bottom panel). $\mathcal{R}_{0,\textrm{LVK}}$ can be marginally matched for $\alpha=10$, $Z_{\textrm{O}/\textrm{H},0}=9.2$ and the lowest $a_{\textrm{MZR}}$ and $\nabla_{\textrm{FMR},0}$.}
   \label{fig:table_alpha10}
\end{figure*}

These results were obtained by considering only the isolated binary merger scenario, that is, we considered that star formation only occurs in the field and completely neglected the effect of dynamics in young, globular, or nuclear star clusters. Since dynamical interactions enhance the BBH merger efficiency, the difference between models and data would only be stronger if these effects were included. The isolated binary merger scenario should either match or underestimate $\mathcal{R}_{0,\textrm{LVK}}$, not overestimate it. 

We recall that all the assumptions and parameters for computing the SFRD$(z, Z)$ were chosen with the aim of trying to minimize the merger rates. This result demonstrates that no reasonable variation in the metallicity relation can reconcile the predicted value of  the local BBH merger rate density with that inferred from LVK data. Our conclusion is that only some modifications in the treatment of single stellar and binary evolution can improve the agreement between the predicted and observed BBH merger rate density. As for single stellar evolution, a possibility is to change the wind treatment used in population synthesis codes, which might decrease the threshold metallicity at which $\eta(Z)$ drops \citep{vanson2025}. Several recent theoretical and observational works have shown that massive stars can develop optically thick winds \citep{Vink2011, Bestenlehner2014, Bestenlehner2020, Brands2022, Brands2025, Sabhahit2022, Sabhahit2023, Romagnolo2024,GormazMatamala2025,Simonato2025, Shepherd2025, Boco2025, Romagnolo2026, Torniamenti2026} even at relatively low metallicities $\sim Z_\odot/5$ \citep{Boco2025}, where mass loss can reach values as high as $10^{-3}\,M_\odot\,\textrm{yr}^{-1}$. For binary evolution, a first possibility is to increase the strength of supernova natal kicks \citep[see discussion in Section \ref{sec:discussion_comparison} and Figure 7 in][]{Sgalletta2025}. Other options include a drastic update of the description of stable mass transfer \citep{Ge2020}, of the common-envelope evolutionary phase \citep{Olejak2021}, and the treatment of tides.

\subsection{The role of $\alpha_{\textrm{CE}}$}\label{sec:results_alpha}
The results we showed in the previous section were obtained assuming a common-envelope efficiency parameter $\alpha_{\textrm{CE}}=1$. Different values of $\alpha{}$ can change the BBH merger rate density as they alter the efficiency of common-envelope ejection. A smaller value of $\alpha{}$ leads to more stellar systems that merge during the common-envelope phase, which in turn reduces the number of available BBHs. A higher value of $\alpha{}$ instaed leads to less shrinking of the binaries and produces fewer BBH systems with a short initial orbital period.

Figure \ref{fig:table_alpha10} shows the results of our analysis for $\alpha_{\textrm{CE}}=0.1$ and $\alpha_{\textrm{CE}}=10$, taken as two extreme values. The predicted merger rate density is more compatible with the values inferred by LVK data for both cases. In particular, the $\alpha=10$ case shows a stronger dependence on metallicity and can drastically reduce the merger rate for the highest-metallicity cases. Even with such an extreme $\alpha_{\textrm{CE}}$ value, however, the most optimistic case yields $\mathcal{R}_0/\mathcal{R}_{0,\textrm{LVK}}\sim3$. We conclude that variations in the common-envelope efficiency are not enough to reconcile the predicted and observed rates \citep{Sgalletta2025}.

\section{Discussion}\label{sec:discussion}
\subsection{Reversing the problem: The metallicity that reconciles the local BBH merger rates}
\begin{figure*}
    \centering
    \includegraphics[width=1.0\linewidth]{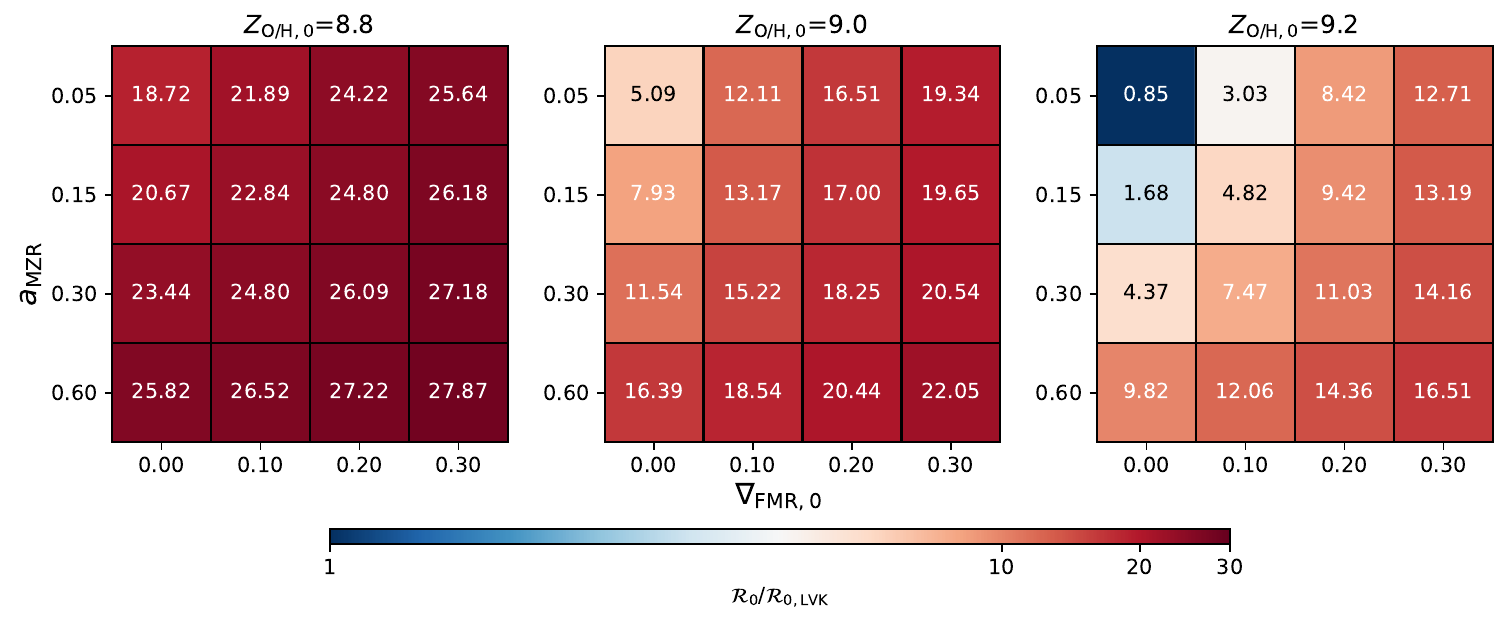}
    \caption{Same as Figure \ref{fig:table}, but for an extended range of parameters. $\mathcal{R}_{0,\textrm{LVK}}$ can be marginally matched only for unrealistic values of $Z_{\textrm{O}/\textrm{H},0}=9.2$, $\nabla_{\textrm{FMR},0}=0$ and $a_{\textrm{MZR}}=0.05, 0.15$.}
    \label{fig:extended_table}
\end{figure*}

\cite{Sgalletta2025} have recently shown that for any reasonable variation in the star-forming galaxy main sequence, the local BBH merger rates cannot be reconciled with observations. Here, we go even further and show that for any reasonable variation in the metallicity relation, the local BBH merger rates are overestimated, even for the most conservative choices for the GSFM and starburst fraction. In this Section, we address the following question: what should the values of the parameters of the metallicity relations (FMR and MZR) be in order to match $\mathcal{R}_{0,\textrm{LVK}}$? To answer this question we extend the parameter space considered, exploring regions with $a_{\textrm{MZR}}=0.05$ and $\nabla_{\textrm{FMR},0}=0, 0.1$. These values are quite unrealistic because they yield metallicities that are far higher than any observationally derived metallicity relation. Figure \ref{fig:extended_table} shows a new matrix with the results of this approach. First of all, even with these extreme variations, only the case with $Z_{\textrm{O}/\textrm{H},0}=9.2$, $\nabla_{\textrm{FMR},0}=0$, and $a_{\textrm{MZR}}=0.05, 0.15$ leads to a reasonable agreement with the observed BBH merger rate.

\begin{figure}
    \centering
    \includegraphics[width=1.0\linewidth]{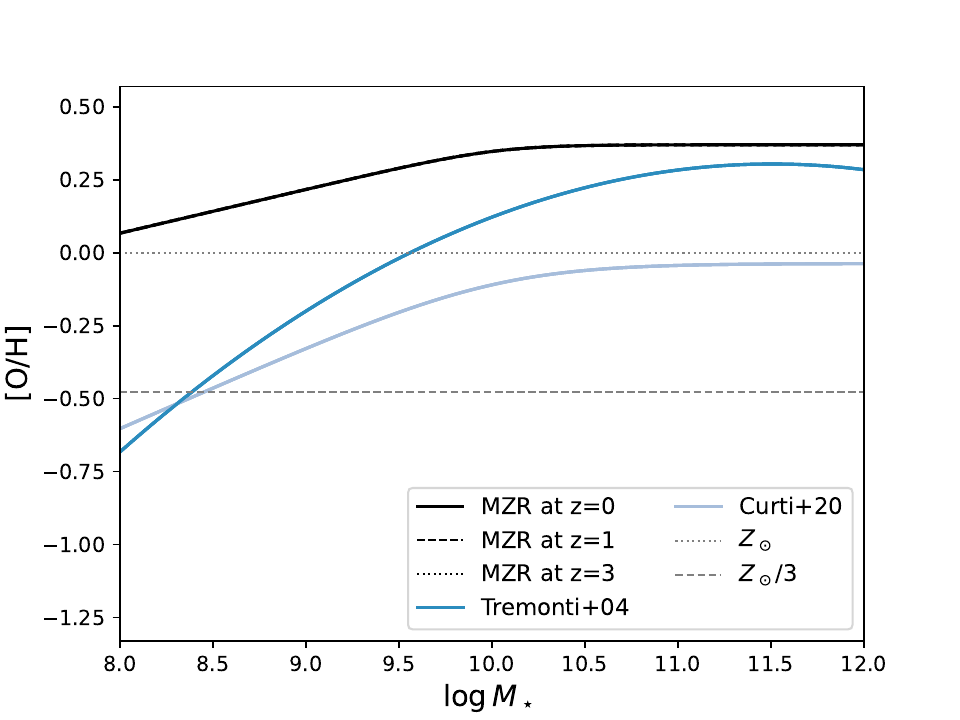}
    \caption{MZR for $Z_{\textrm{O}/\textrm{H},0}=9.2$, $\nabla_{\textrm{FMR},0}=0$, and $a_{\textrm{MZR}}=0.15$. The solid, dashed, and dotted blue lines show the MZR at $z=0,\,1,$ and $3$, but they are superimposed since the MZR does not evolve with redshift for $\nabla_{\textrm{FMR},0}=0$. The two blue lines are classical $z=0$ MZR from the literature: \cite{Tremonti2004} in dark blue, and \cite{Curti2020} in steel blue. The dashed and dotted horizontal gray lines represent $Z_\odot/3$ and $Z_\odot$. Unrealistically high values of metallicity are needed to reproduce the local BBH merger rate density, especially for low-mass galaxies.}
    \label{fig:metallicity_relation_high}
\end{figure}

The metallicity relation associated with the $\nabla_{\textrm{FMR},0}=0$, $a_{\textrm{MZR}}=0.15$ case is shown in Figure \ref{fig:metallicity_relation_high} in terms of $[\textrm{O}/\textrm{H}]$. Matching the local BBH merger rate density would require extremely high metallicities, both in terms of normalization, and
in terms of the steepness of the relation, implying that galaxies with mass $M_\star=10^8\,\textrm{M}_\odot$ should have an average oxygen abundance around the solar value. This value is higher than what found in observations, which indicate metallicities in the range $\sim 0.2-0.3\,\textrm{Z}_\odot$ for $M_\star=10^8\,\textrm{M}_\odot$ galaxies at $z=0$. Even more importantly, the absence of a correlation with SFR ($\nabla_{\textrm{FMR},0}=0$) implies that metallicity should not evolve with redshift, meaning that these high metallicity values are the same at higher redshift.

\subsection{The effect of the merger efficiency and delay-time distribution}
Here, we explore how the merger efficiency and the delay time distribution affect $\mathcal{R}_0$. Specifically, we investigate which efficiencies and delay times can reconcile the predictions of the local BBH merger rate with observations. 

To this end, we vary the merger efficiency and the delay-time distribution separately while keeping the other quantity fixed. Specifically, we lower the merger efficiency while keeping the delay-time distribution fixed, and we steepen the delay time distribution while keeping $\eta(Z)$ fixed.

As the merger efficiency remains nearly constant for $Z<Z_\odot/3$ and declines sharply at higher metallicity, it acts as a multiplicative factor in the computation of the merger rates. Consequently, rescaling the low-metallicity merger efficiency by a given factor results in an equivalent rescaling of the predicted merger rate density. Figure \ref{fig:table} shows that to match $\mathcal{R}_{0,\textrm{LVK}}$, a reduction in $\eta(Z)$ by a factor $\gtrsim 10$ is required. 

An alternative approach is to further decrease the metallicity threshold above which the merger efficiency drops significantly, thereby excluding a larger fraction of the SFRD from contributing to BBH mergers. The redshift evolution of the BBH merger rates and the observed properties of BBHs can distinguish between these two alternatives (see Bosi et al. 2026, in prep.). 

\begin{figure*}
    \centering
    \includegraphics[width=0.9\textwidth]{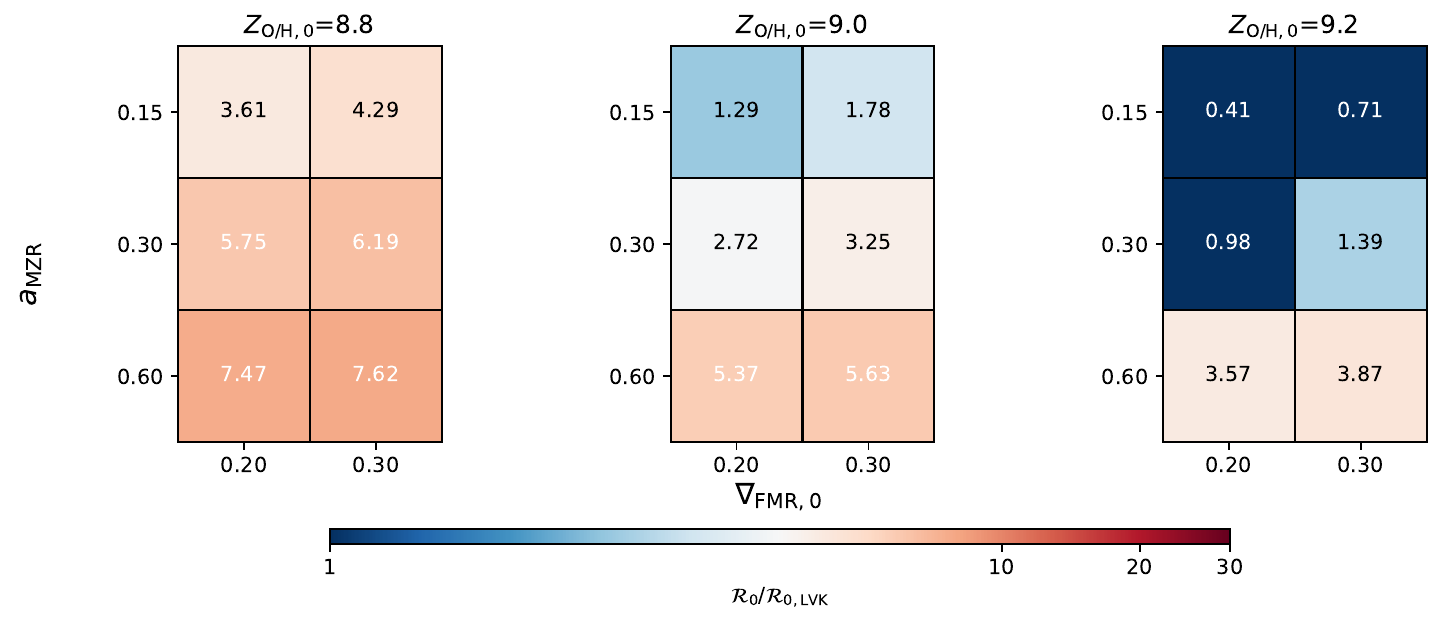}
    \caption{Same as Figure \ref{fig:table}, but for $\textrm{d}p/\textrm{d}t_d\propto t_d^{-1.5}$. $\mathcal{R}_{0,\textrm{LVK}}$ can be matched for a large set of metallicity parameters.}
    \label{fig:table_delay}
\end{figure*}

For the delay-time distribution, we assume a power-law scaling $\propto t_d^{-\alpha}$ and steepen the exponent $\alpha$ until an agreement with $\mathcal{R}_{0,\textrm{LVK}}$ is obtained. The steepening of the delay-time distribution rapidly decreases the local merger rate density \citep{Smith2026a}. Figure \ref{fig:table_delay} shows the results for $\textrm{d}p/\textrm{d}t_d\propto t_d^{-1.5}$. The agreement with the observed value is now much easier to reach, especially for the assumption $Z_{\textrm{O}/\textrm{H},0}=9.2$. In some cases, the choice of $Z_{\textrm{O}/\textrm{H},0}=9.2$ even leads to underestimated rates. 

Realistic metallicity relations with $Z_{\textrm{O}/\textrm{H},0}=9.0$, $a_{\textrm{MZR}}=0.15$, $\nabla_{\textrm{FMR},0}=0.2-0.3$ can now reproduce the local merger rate density within a factor $\lesssim 2$. This happens because a steepening of the delay time distribution implies that more BBHs merge sooner after their formation and the contribution to the local BBH merger rate from objects that formed at high redshift is reduced. This is important because a substantial fraction of the local BBH merger rate is contributed by objects that formed at $z\sim 2$, where the cosmic SFR of the Universe peaks.

However, we caution that the results of this section are meant just as a toy model of the behavior that binary population simulations should exhibit in order to reconcile the predicted $\mathcal{R}_0$ with observations, not as a solution to the problem. The approach of separately varying the merger efficiency or delay time is indeed intrinsically inconsistent. Since the merger efficiency is defined as the number of BBH mergers in a Hubble time, $t_H$, divided by the total mass of the stellar population, its value is strongly connected to the shape of the delay-time distribution,

\begin{equation}
    \eta(Z)\equiv\frac{1}{M_{\textrm{pop}}}\,\int_{t_{d,min}}^{t_H}\textrm{d}t_d\,\frac{\textrm{d}N}{\textrm{d}t_d}(t_d|Z).
\end{equation}

If the number of BBHs formed is fixed, a steeper delay-time distribution results in a higher merger efficiency $\eta(Z)$ because more systems have delay times $t_d<t_H$ and therefore merge within a Hubble time. Conversely, a shallower delay-time distribution leads to a lower value of $\eta(Z)$ because fewer systems satisfy $t_d>t_H$. 

The pronounced decline in the merger efficiency at high metallicities can be explained almost entirely by the corresponding change in the shape of the delay-time distribution, which becomes shallower with increasing metallicity. This behavior was clearly illustrated by \citet{Iorio2023} (see their Fig. 17), who showed that the BBH formation efficiency, $\eta_{\textrm{form}}$, decreases only by a factor of $\sim 2$ across the full metallicity range. In contrast, the merger efficiency, $\eta$, drops by more than three orders of magnitude at $Z\gtrsim Z_\odot/3$. This difference can be entirely attributed to the behavior of the delay-time distribution.

This interplay between merger efficiency and delay-time distribution produces a compensation effect in the computation of the local BBH merger rates: shorter delay-time distributions imply a higher merger efficiency, but fewer systems with long delay times, thereby suppressing the contribution of BBHs that formed at cosmic noon ($z\sim{2}$) to the local merger rate. Conversely, a shallower delay-time distribution decreases the merger efficiency, but enhances the contribution to the local merger rate of BBHs that formed at the peak of the star formation activity (see, e.g., Figure 11 of \cite{Santoliquido2021}).

\subsection{Comparison with previous works}\label{sec:discussion_comparison}

Several studies estimated the BBH merger rate density locally and at high redshift \citep{OShaughnessy2010, Dominik2013, Belczynski2016, Lamberts2016, Mapelli2017, Oshaughnessy2017, Lamberts2018, Mapelli2018, Kruckow2018, Bray2018, Cao2018, Elbert2018, Li2018, Boco2019, Artale2019, Neijssel2019, Tang2020, Santoliquido2020, Boco2021, Santoliquido2021, Mapelli2021, Olejak2021, Broekgaarden2021, Broekgaarden2022, Santoliquido2022, Ghodla2022, Karathanasis2023, Briel2023, Iorio2023, Romagnolo2023,Romagnolo2025,Sgalletta2025, Levina2026, Smith2026a, Padhyegurjar2026a, Padhyegurjar2026b}. Since the computation involves modeling several physical processes, many assumptions, choices, and parameters, it is a complex task to thoroughly compare all the results. However, since some of the model variations did find a local BBH merger rate within or below the $90\%$ credible interval inferred by LVK, it is instructive to analyze the main reasons for these low merger rates in some models. We use the review by \cite{Mandel2022} and \cite{Broekgaarden2026b} as a guidance through the $\mathcal{R}_0$ estimations of different works (see their Figure 3 and Figure 1, respectively). The studies reported there investigated different model variations both for the stellar and binary evolution term and for the metallicity-dependent SFRD, obtaining a wide range of $\mathcal{R}_0$ values. Most of the works predict $\mathcal{R}_0\gtrsim 100\,\textrm{Gpc}^{-3}\,\textrm{yr}^{-1}$, but some of them feature model variations extending to $\mathcal{R}_0\lesssim 20\,\textrm{Gpc}^{-3}\,\textrm{yr}^{-1}$. Here, we considered only the latter and analyze only the model variations leading to such low rates with the aim to find common assumptions and features.

The reasons for the low BBH merger rates predicted by some models can be various \citep{Broekgaarden2026b} and concern both the treatment of the metal-dependent SFRD and stellar and binary evolution physics.

\subsubsection{Metal-dependent SFRD}
For the metal-dependent SFRD, low BBH merger rates are obtained by models with no or low scatter around the metallicity relation \citep{Dominik2013, Santoliquido2020, Santoliquido2021, Mapelli2021, Iorio2023, ArcaSedda2026}. In particular, most of these models use a cosmic averaged metallicity, that is, an average metallicity at each redshift, with a fixed scatter, usually assumed to follow a Gaussian $\log{Z}$ distribution with $\sigma\sim 0.2\,\textrm{dex}$, underestimating low-metallicity tails due to low-mass galaxies. This is shown in Figure 4 of \cite{Santoliquido2022}, which compares the metallicity dispersion in models following an FMR or MZR and cosmic-averaged models. Also \cite{Tang2020} showed that $\mathcal{R}_0\lesssim 20\,\textrm{Gpc}^{-3}\,\textrm{yr}^{-1}$ can only be obtained for $\beta\gtrsim 4$ (see their Eq. 4), corresponding to very low scatter in the cosmic-averaged metallicity. Some of the models also assumed fixed \citep{Kruckow2018} and sometimes very high \citep{OShaughnessy2010, Bray2018} metallicities, with no redshift evolution. 

A useful overview of how metallicity affects the local BBH merger rate density can be found in \cite{Broekgaarden2022}. Most of the models reported there tend to produce $\mathcal{R}_0> 20\,\textrm{Gpc}^{-3}\,\textrm{yr}^{-1}$. The only models matching $\mathcal{R}_0\sim 20\,\textrm{Gpc}^{-3}\,\textrm{yr}^{-1}$ are those with very high average metallicities. This is clearly shown in their Figs. A2 and in Fig. B1 \citep{Broekgaarden2021}, where their model with the lowest merger rates is shown to have an average metallicity $Z\sim Z_\odot/2$ up to $z\sim 3$. Finally, in most of the works mentioned above, oxygen abundances were used to derive metallicity, which further underestimated the merger rates.

In this work, we have shown that using an observationally derived SFRD with an extended low-metallicity tail and considering iron instead of oxygen abundance greatly increase the derived $\mathcal{R}_0$ value. In \citep{Boco2026b}, we have demonstrated that this is valid for many different population synthesis codes, under different physical assumptions and parameter choices.

\subsubsection{Stellar and binary evolution}
As for stellar and binary evolution, pinpointing the exact physical mechanism that can reconcile theory with observations is challenging, given the variety of physical assumptions in different studies and the high dimensionality of the parameter space. However, many of the models from the literature that predict the lowest rates involve either high natal kicks or changes in the mass-transfer and common-envelope physics.

For the natal kicks, only model variations using the kick prescriptions of \cite{Hobbs2005}, with $\sigma=265\,\textrm{km}\,\textrm{s}^{-1}$ or higher \citep[$\sigma_{\textrm{rms}}^\textrm{1D}=750$ km s$^{-1}$, ][]{Boesky2024b}, and without fallback, produce low $\mathcal{R}_0$ values. This is valid for \textsc{bse} \citep{Mapelli2017}, \textsc{StarTrack} \citep{Chruslinska2019a, Belczynski2020}, \textsc{compas} \citep{Boesky2024b}, and \textsc{sevn} \citep{Sgalletta2025}, while \textsc{bpass} produces high local rates even with such strong kicks \citep{Eldridge2019, Tang2020}. While the \cite{Hobbs2005} kicks are calibrated on neutron star observations, though with the correction introduced by \cite{Disberg2025}, constraining BH natal kicks is more challenging \citep{Repetto2012, Repetto2017, Atri2019}. Historically, BH kicks have generally been assumed to be weaker than neutron star kicks because of fallback during the supernova explosion \citep{Mirabel2003, Belczynski2010, Giacobbo2018, Neijssel2019}. Here, we show that neutron star-like kicks can help alleviate the overestimation of BBH merger rates. The next \textit{Gaia} data release will provide useful constraints on the kicks: if kicks are modulated by fallback, there should be a large population of dormant BHs in long-period binaries that the next \textit{Gaia} data release should observe \citep{Mapelli2026}. A non-detection of such a population by \textit{Gaia} would favor stronger BH natal kicks, similar to those inferred for Galactic neutron stars.

Regarding mass transfer, models assuming exclusively stable mass transfer \citep{vanson2022b} can substantially reduce the predicted merger rates \citep[see also][]{Sgalletta2025}. Moreover, the adoption of more restrictive criteria for mass-transfer instability \citep{Ge2020, Olejak2021}, more conservative assumptions about common-envelope survival \citep{Romagnolo2025}, or a modified treatment of rejuvenation during mass transfer \citep{Ghodla2022, Briel2023} can lead to merger rates that are consistent with LVK observations even for a metallicity distribution with an extended low-metallicity tail, $\sigma_Z=0.4-0.5$.

\begin{figure*}
    \centering
    \sidecaption
    \includegraphics[width=0.9\textwidth]{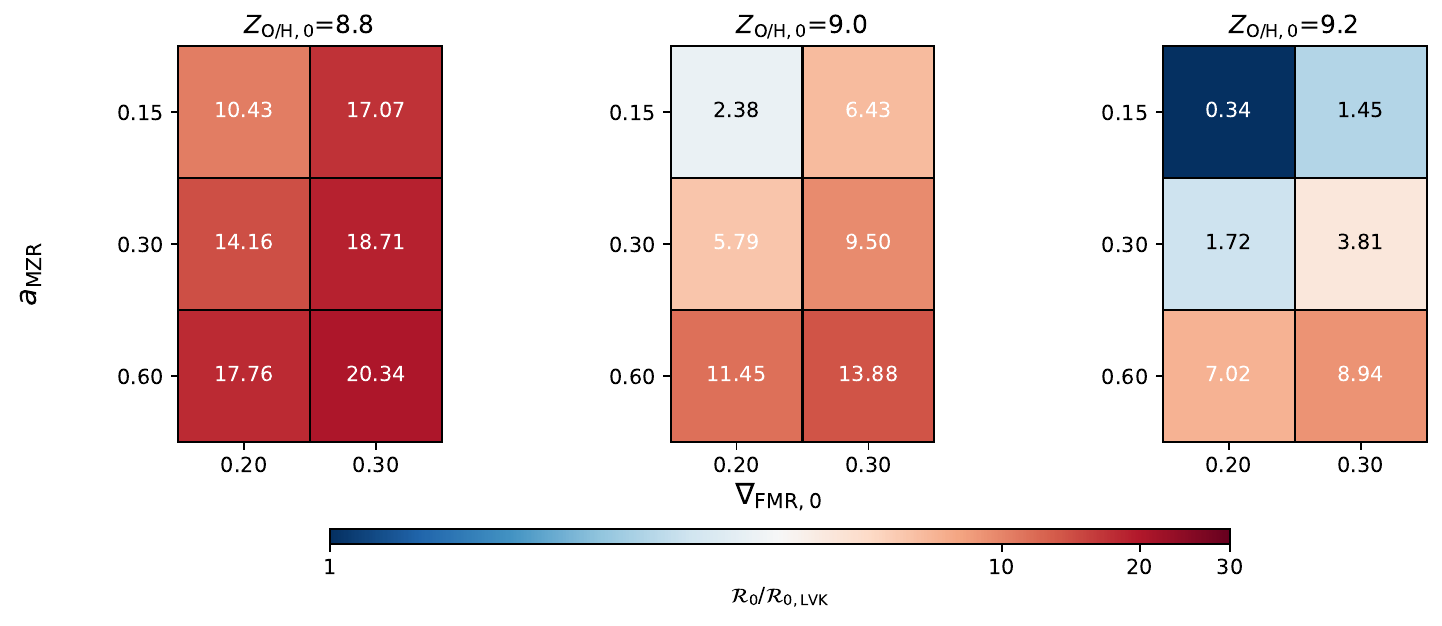}
    \caption{Same as Figure \ref{fig:table}, but using oxygen abundances.}
    \label{fig:table_oxygen}
\end{figure*}

Despite the diﬀerent modelization present in the works, both on the galaxy evolution side and
on the binary evolution one, we conclude that these results agree with the main message of this paper: when an observationally derived SFRD is adopted and the iron abundance is considered, the predicted local BBH merger rates can be reconciled with observations only via a reduction of the BBH merger efficiency. Possible ways to achieve this are an increase in the strength of natal kicks or a different treatment of the mass transfer and common-envelope evolution in binary population synthesis codes.

\subsection{Impact of the iron abundance}\label{sec:iron}
Finally, we investigate the impact on $\mathcal{R}_0$ of adopting the iron abundance rather than oxygen. In Figure \ref{fig:table_oxygen} we present results obtained using oxygen abundances, without the correction for $[\textrm{O}/\textrm{Fe}]$. Even in this case, most models still overestimate $\mathcal{R}_0$. For some configurations with $Z_{\textrm{O}/\textrm{H},0}=9.2$, however, the observed local merger rate density can be reproduced within a factor of $\sim 2$. We conclude that the choice of the iron rather than oxygen abundance has a substantial impact, significantly exacerbating the tension between models and observations. In particular, variations yielding the highest metallicities are more affected by the use of iron abundances. This is because the iron abundance introduces the necessary scatter to allow a non-negligible fraction of star formation to occur at low metallicities ($Z<Z_\odot/3$), thereby increasing the predicted BBH merger rate density.

\section{Summary}\label{sec:conclusions}
We estimated the BBH merger rate density from the isolated binary-evolution channel using empirical relations to compute the metallicity-dependent SFRD and combining it with the outcomes of the binary population synthesis code \textsc{sevn}. For the first time in an empirical approach, we have used iron abundances rather than oxygen abundances to perform the computation. Our main findings are listed below.
\begin{itemize}
    \item The predicted local BBH merger rate density is higher than the observed $\mathcal{R}_{0,\textrm{LVK}}\sim 14-26\,\textrm{Gpc}^{-3}\,\textrm{yr}^{-1}$ for any reasonable variation in the fundamental metallicity relation by a factor $> 10$. This is true even when the metallicity-dependent SFRD is built with the aim of minimizing local BBH merger rates.
    \item The predicted and observed $\mathcal{R}_0$ value can only be reconciled with extremely high metallicities, $\sim Z_\odot$, even for low-mass galaxies $\sim 10^8\,\textrm{M}_\odot$ at high redshift.  
    \item Changing the common-envelope efficiency parameter $\alpha_{\textrm{CE}}$ alleviates the problem, but even extreme values of $\alpha_{\textrm{CE}}=10$ yield $\mathcal{R}_0/\mathcal{R}_{0,\textrm{LVK}}\gtrsim 3$ only for the most optimistic metallicity variations.
    \item A reduced merger efficiency by a factor of $\sim 10$ or a steepening of the delay-time distribution to $\textrm{d}p/\textrm{d}t_d\propto t_d^{-1.5}$ is needed in order to reconcile the predicted and observed local merger rate density.
    \item Using the iron abundance instead of the oxygen abundance \citetalias{Chruslinska2025} has a large impact on the estimated merger rate density and worsens the tension with data by a factor $\sim 2-10$, depending on the metallicity variation.
\end{itemize}

Overall, our results show that variations in the metallicity-dependent SFRD within the limits allowed by observations are not enough to reconcile the predicted and observed local BBH merger rate density. We advocate modifications to the treatment of stellar and binary evolution to obtain successful predictions. Some possibilities include higher natal kicks, a revised treatment of the mass transfer and common-envelope evolution, and/or a revised treatment of massive star winds. In future works, we will explore some of these options and their effects on the merger rate and on the properties of binary compact objects.

\begin{acknowledgements}
We thank the anonymous referee for helpful comments and suggestions. LB thanks Stefano Torniamenti for useful discussions. LB acknowledges support by the Deutsche Forschungsgemeinschaft (DFG, German Research Foundation) in the form of a Walter Benjamin position -- Projektnummer 555003977. LB, MM and CS acknowledge financial support from the German Excellence Strategy via the Heidelberg Cluster of Excellence (EXC 2181 - 390900948) STRUCTURES. MM acknowledges financial support from the European Research Council for the ERC Consolidator grant DEMOBLACK, under contract no. 770017. CS acknowledges financial support from the Alexander von Humboldt Foundation for the Humboldt Research Fellowship. MB acknowledges that this article was produced while attending the PhD program in PhD in Space Science and Technology at the University of Trento, Cycle XXXIX, with the support of a scholarship financed by the Ministerial Decree no. 118 of 2nd March 2023, based on the NRRP - funded by the European Union - NextGenerationEU - Mission 4 "Education and Research", Component 1 "Enhancement of the offer of educational services: from nurseries to universities” - Investment 4.1 “Extension of the number of research doctorates and innovative doctorates for public administration and cultural heritage” - CUP E66E23000110001 and support by the Italian grant Project SPACE-IT-UP by the Italian Space Agency and Ministry of University and Research, Contract Number 2024-5-E.0.
This research made use of \textsc{NumPy} \citep{Harris20}, \textsc{SciPy} \citep{SciPy2020}, and \textsc{Matplotlib} \citep{Hunter2007}. We used \textsc{sevn} (\href{https://gitlab.com/sevncodes/sevn}{https://gitlab.com/sevncodes/sevn}) to generate our BBHs catalogs \citep{Spera2019, Mapelli2020b, Iorio2023}.

\end{acknowledgements}

\begingroup
\small
\bibliographystyle{aa}
\bibliography{references}

@ARTICLE{Klencki2018,
       author = {{Klencki}, J. and {Moe}, M. and {Gladysz}, W. and {Chruslinska}, M. and {Holz}, D.~E. and {Belczynski}, K.},
        title = "{Impact of inter-correlated initial binary parameters on double black hole and neutron star mergers}",
      journal = {\aap},
         year = 2018,
        month = nov,
       volume = {619},
          eid = {A77},
        pages = {A77},
          doi = {10.1051/0004-6361/201833025},
archivePrefix = {arXiv},
       eprint = {1808.07889},
 primaryClass = {astro-ph.HE},
       adsurl = {https://ui.adsabs.harvard.edu/abs/2018A&A...619A..77K}
}

@ARTICLE{Harris20,
       author = {{Harris}, Charles R. and {Millman}, K. Jarrod and {van der Walt}, St{\'e}fan J. and {Gommers}, Ralf and {Virtanen}, Pauli and {Cournapeau}, David and {Wieser}, Eric and {Taylor}, Julian and {Berg}, Sebastian and {Smith}, Nathaniel J. and {Kern}, Robert and {Picus}, Matti and {Hoyer}, Stephan and {van Kerkwijk}, Marten H. and {Brett}, Matthew and {Haldane}, Allan and {del R{\'\i}o}, Jaime Fern{\'a}ndez and {Wiebe}, Mark and {Peterson}, Pearu and {G{\'e}rard-Marchant}, Pierre and {Sheppard}, Kevin and {Reddy}, Tyler and {Weckesser}, Warren and {Abbasi}, Hameer and {Gohlke}, Christoph and {Oliphant}, Travis E.},
        title = "{Array programming with NumPy}",
      journal = {\nat},
         year = 2020,
        month = sep,
       volume = {585},
       number = {7825},
        pages = {357-362},
          doi = {10.1038/s41586-020-2649-2},
archivePrefix = {arXiv},
       eprint = {2006.10256},
 primaryClass = {cs.MS},
       adsurl = {https://ui.adsabs.harvard.edu/abs/2020Natur.585..357H}
}

@ARTICLE{SciPy2020,
       author = {{Virtanen}, Pauli and {Gommers}, Ralf and {Oliphant}, Travis E. and {Haberland}, Matt and {Reddy}, Tyler and {Cournapeau}, David and {Burovski}, Evgeni and {Peterson}, Pearu and {Weckesser}, Warren and {Bright}, Jonathan and {van der Walt}, St{\'e}fan J. and {Brett}, Matthew and {Wilson}, Joshua and {Millman}, K. Jarrod and {Mayorov}, Nikolay and {Nelson}, Andrew R.~J. and {Jones}, Eric and {Kern}, Robert and {Larson}, Eric and {Carey}, C.~J. and {Polat}, {\.I}lhan and {Feng}, Yu and {Moore}, Eric W. and {VanderPlas}, Jake and {Laxalde}, Denis and {Perktold}, Josef and {Cimrman}, Robert and {Henriksen}, Ian and {Quintero}, E.~A. and {Harris}, Charles R. and {Archibald}, Anne M. and {Ribeiro}, Ant{\^o}nio H. and {Pedregosa}, Fabian and {van Mulbregt}, Paul and {SciPy 1. 0 Contributors}},
        title = "{SciPy 1.0: fundamental algorithms for scientific computing in Python}",
      journal = {Nature Methods},
         year = 2020,
        month = feb,
       volume = {17},
        pages = {261-272},
          doi = {10.1038/s41592-019-0686-2},
archivePrefix = {arXiv},
       eprint = {1907.10121},
 primaryClass = {cs.MS},
       adsurl = {https://ui.adsabs.harvard.edu/abs/2020NatMe..17..261V}
}

@ARTICLE{Hunter2007,
       author = {{Hunter}, John D.},
        title = "{Matplotlib: A 2D Graphics Environment}",
      journal = {Computing in Science and Engineering},
         year = 2007,
        month = may,
       volume = {9},
       number = {3},
        pages = {90-95},
          doi = {10.1109/MCSE.2007.55},
       adsurl = {https://ui.adsabs.harvard.edu/abs/2007CSE.....9...90H}
}

@ARTICLE{Spera2019,
       author = {{Spera}, Mario and {Mapelli}, Michela and {Giacobbo}, Nicola and {Trani}, Alessandro A. and {Bressan}, Alessandro and {Costa}, Guglielmo},
        title = "{Merging black hole binaries with the SEVN code}",
      journal = {\mnras},
         year = 2019,
        month = may,
       volume = {485},
       number = {1},
        pages = {889-907},
          doi = {10.1093/mnras/stz359},
archivePrefix = {arXiv},
       eprint = {1809.04605},
 primaryClass = {astro-ph.HE},
       adsurl = {https://ui.adsabs.harvard.edu/abs/2019MNRAS.485..889S}
}

@ARTICLE{Mapelli2020b,
       author = {{Mapelli}, Michela and {Spera}, Mario and {Montanari}, Enrico and {Limongi}, Marco and {Chieffi}, Alessandro and {Giacobbo}, Nicola and {Bressan}, Alessandro and {Bouffanais}, Yann},
        title = "{Impact of the Rotation and Compactness of Progenitors on the Mass of Black Holes}",
      journal = {\apj},
         year = 2020,
        month = jan,
       volume = {888},
       number = {2},
          eid = {76},
        pages = {76},
          doi = {10.3847/1538-4357/ab584d},
archivePrefix = {arXiv},
       eprint = {1909.01371},
 primaryClass = {astro-ph.HE},
       adsurl = {https://ui.adsabs.harvard.edu/abs/2020ApJ...888...76M}
}

@ARTICLE{Iorio2023,
       author = {{Iorio}, Giuliano and {Mapelli}, Michela and {Costa}, Guglielmo and {Spera}, Mario and {Escobar}, Gast{\'o}n J. and {Sgalletta}, Cecilia and {Trani}, Alessandro A. and {Korb}, Erika and {Santoliquido}, Filippo and {Dall'Amico}, Marco and {Gaspari}, Nicola and {Bressan}, Alessandro},
        title = "{Compact object mergers: exploring uncertainties from stellar and binary evolution with SEVN}",
      journal = {\mnras},
         year = 2023,
        month = sep,
       volume = {524},
       number = {1},
        pages = {426-470},
          doi = {10.1093/mnras/stad1630},
archivePrefix = {arXiv},
       eprint = {2211.11774},
 primaryClass = {astro-ph.HE},
       adsurl = {https://ui.adsabs.harvard.edu/abs/2023MNRAS.524..426I}
}

@ARTICLE{Santoliquido2021,
       author = {{Santoliquido}, Filippo and {Mapelli}, Michela and {Giacobbo}, Nicola and {Bouffanais}, Yann and {Artale}, M. Celeste},
        title = "{The cosmic merger rate density of compact objects: impact of star formation, metallicity, initial mass function, and binary evolution}",
      journal = {\mnras},
         year = 2021,
        month = apr,
       volume = {502},
       number = {4},
        pages = {4877-4889},
          doi = {10.1093/mnras/stab280},
archivePrefix = {arXiv},
       eprint = {2009.03911},
 primaryClass = {astro-ph.HE},
       adsurl = {https://ui.adsabs.harvard.edu/abs/2021MNRAS.502.4877S}
}

@ARTICLE{Sabhahit2023,
       author = {{Sabhahit}, Gautham N. and {Vink}, Jorick S. and {Sander}, Andreas A.~C. and {Higgins}, Erin R.},
        title = "{Very massive stars and pair-instability supernovae: mass-loss framework for low metallicity}",
      journal = {\mnras},
         year = 2023,
        month = sep,
       volume = {524},
       number = {1},
        pages = {1529-1546},
          doi = {10.1093/mnras/stad1888},
archivePrefix = {arXiv},
       eprint = {2306.11785},
 primaryClass = {astro-ph.SR},
       adsurl = {https://ui.adsabs.harvard.edu/abs/2023MNRAS.524.1529S}
}

@ARTICLE{Sabhahit2022,
       author = {{Sabhahit}, Gautham N. and {Vink}, Jorick S. and {Higgins}, Erin R. and {Sander}, Andreas A.~C.},
        title = "{Mass-loss implementation and temperature evolution of very massive stars}",
      journal = {\mnras},
         year = 2022,
        month = aug,
       volume = {514},
       number = {3},
        pages = {3736-3753},
          doi = {10.1093/mnras/stac1410},
archivePrefix = {arXiv},
       eprint = {2205.09125},
 primaryClass = {astro-ph.SR},
       adsurl = {https://ui.adsabs.harvard.edu/abs/2022MNRAS.514.3736S}
}

@ARTICLE{Bestenlehner2014,
       author = {{Bestenlehner}, J.~M. and {Gr{\"a}fener}, G. and {Vink}, J.~S. and {Najarro}, F. and {de Koter}, A. and {Sana}, H. and {Evans}, C.~J. and {Crowther}, P.~A. and {H{\'e}nault-Brunet}, V. and {Herrero}, A. and {Langer}, N. and {Schneider}, F.~R.~N. and {Sim{\'o}n-D{\'\i}az}, S. and {Taylor}, W.~D. and {Walborn}, N.~R.},
        title = "{The VLT-FLAMES Tarantula Survey. XVII. Physical and wind properties of massive stars at the top of the main sequence}",
      journal = {\aap},
         year = 2014,
        month = oct,
       volume = {570},
          eid = {A38},
        pages = {A38},
          doi = {10.1051/0004-6361/201423643},
archivePrefix = {arXiv},
       eprint = {1407.1837},
 primaryClass = {astro-ph.SR},
       adsurl = {https://ui.adsabs.harvard.edu/abs/2014A&A...570A..38B}
}

@ARTICLE{Vink2011,
       author = {{Vink}, Jorick S. and {Muijres}, L.~E. and {Anthonisse}, B. and {de Koter}, A. and {Gr{\"a}fener}, G. and {Langer}, N.},
        title = "{Wind modelling of very massive stars up to 300 solar masses}",
      journal = {\aap},
         year = 2011,
        month = jul,
       volume = {531},
          eid = {A132},
        pages = {A132},
          doi = {10.1051/0004-6361/201116614},
archivePrefix = {arXiv},
       eprint = {1105.0556},
 primaryClass = {astro-ph.SR},
       adsurl = {https://ui.adsabs.harvard.edu/abs/2011A&A...531A.132V}
}

@ARTICLE{Simonato2025,
       author = {{Simonato}, Filippo and {Torniamenti}, Stefano and {Mapelli}, Michela and {Iorio}, Giuliano and {Boco}, Lumen and {De Domenico-Langer}, Franca and {Sgalletta}, Cecilia},
        title = "{Impact of stellar winds on the pair-instability supernova rate}",
      journal = {\aap},
         year = 2025,
        month = nov,
       volume = {703},
          eid = {A215},
        pages = {A215},
          doi = {10.1051/0004-6361/202555490},
archivePrefix = {arXiv},
       eprint = {2505.07959},
 primaryClass = {astro-ph.HE},
       adsurl = {https://ui.adsabs.harvard.edu/abs/2025A&A...703A.215S}
}

@ARTICLE{Abac2026,
       author = {{Abac}, A.~G. and {Abouelfettouh}, I. and {Acernese}, F. and {Ackley}, K. and {Adamcewicz}, C. and {Adhicary}, S. and {Adhikari}, D. and {Adhikari}, N. and {Adhikari}, R.~X. and {Adkins}, V.~K. and {Afroz}, S. and {Agarwal}, D. and {Agathos}, M. and {Aghaei Abchouyeh}, M. and {Aguiar}, O.~D. and {Ahmadzadeh}, S. and {Aiello}, L. and {Ain}, A. and {Ajith}, P. and {Akutsu}, T. and {Albanesi}, S. and {Alfaidi}, R.~A. and {Al-Jodah}, A. and {All{\'e}n{\'e}}, C. and {Allocca}, A. and {Al-Shammari}, S. and {Altin}, P.~A. and {Alvarez-Lopez}, S. and {Amarasinghe}, O. and {Amato}, A. and {Amra}, C. and {Ananyeva}, A. and {Anderson}, S.~B. and {Anderson}, W.~G. and {Andia}, M. and {Ando}, M. and {Andrade}, T. and {Andr{\'e}s-Carcasona}, M. and {Andri{\'c}}, T. and {Anglin}, J. and {Ansoldi}, S. and {Antelis}, J.~M. and {Antier}, S. and {Aoumi}, M. and {Appavuravther}, E.~Z. and {Appert}, S. and {Apple}, S.~K. and {Arai}, K. and {Araya}, A. and {Araya}, M.~C. and {Arca Sedda}, M. and {Areeda}, J.~S. and {Argianas}, L. and {Aritomi}, N. and {Armato}, F. and {Armstrong}, S. and {Arnaud}, N. and {Arogeti}, M. and {Aronson}, S.~M. and {Arun}, K.~G. and {Ashton}, G. and {Aso}, Y. and {Assiduo}, M. and {Assis de Souza Melo}, S. and {Aston}, S.~M. and {Astone}, P. and {Attadio}, F. and {Aubin}, F. and {Aultoneal}, K. and {Avallone}, G. and {Babak}, S. and {Badaracco}, F. and {Badger}, C. and {Bae}, S. and {Bagnasco}, S. and {Bagui}, E. and {Baiotti}, L. and {Bajpai}, R. and {Baka}, T. and {Baker}, T. and {Ball}, M. and {Ballardin}, G. and {Ballmer}, S.~W. and {Banagiri}, S. and {Banerjee}, B. and {Bankar}, D. and {Baptiste}, T.~M. and {Baral}, P. and {Barayoga}, J.~C. and {Barish}, B.~C. and {Barker}, D. and {Barman}, N. and {Barneo}, P. and {Barone}, F. and {Barr}, B. and {Barsotti}, L. and {Barsuglia}, M. and {Barta}, D. and {Bartoletti}, A.~M. and {Barton}, M.~A. and {Bartos}, I. and {Basak}, S. and {Basalaev}, A. and {Bassiri}, R. and {Basti}, A. and {Bates}, D.~E. and {Bawaj}, M. and {Baxi}, P. and {Bayley}, J.~C. and {Baylor}, A.~C. and {Baynard}, II, P.~A. and {Bazzan}, M. and {Bedakihale}, V.~M. and {Beirnaert}, F. and {Bejger}, M. and {Belardinelli}, D. and {Bell}, A.~S. and {Bellie}, D.~S. and {Bellizzi}, L. and {Beltran-Martinez}, D. and {Benoit}, W. and {Bentara}, I. and {Bentley}, J.~D. and {Ben Yaala}, M. and {Bera}, S. and {Bergamin}, F. and {Berger}, B.~K. and {Bernuzzi}, S. and {Beroiz}, M. and {Berry}, C.~P.~L. and {Bersanetti}, D. and {Bertolini}, A. and {Betzwieser}, J. and {Beveridge}, D. and {Bevilacqua}, G. and {Bevins}, N. and {Bhandare}, R. and {Bhatt}, R. and {Bhattacharjee}, D. and {Bhaumik}, S. and {Bhowmick}, S. and {Biancalana}, V. and {Bianchi}, A. and {Bilenko}, I.~A. and {Billingsley}, G. and {Binetti}, A. and {Bini}, S. and {Binu}, C. and {Birnholtz}, O. and {Biscoveanu}, S. and {Bisht}, A. and {Bitossi}, M. and {Bizouard}, M.-A. and {Blaber}, S. and {Blackburn}, J.~K. and {Blagg}, L.~A. and {Blair}, C.~D. and {Blair}, D.~G. and {Bobba}, F. and {Bode}, N. and {Boileau}, G. and {Boldrini}, M. and {Bolingbroke}, G.~N. and {Bolliand}, A. and {Bonavena}, L.~D. and {Bondarescu}, R. and {Bondu}, F. and {Bonilla}, E. and {Bonilla}, M.~S. and {Bonino}, A. and {Bonnand}, R. and {Booker}, P. and {Borchers}, A. and {Borhanian}, S. and {Boschi}, V. and {Bose}, S. and {Bossilkov}, V. and {Boudon}, A. and {Bozzi}, A. and {Bradaschia}, C. and {Brady}, P.~R. and {Branch}, A. and {Branchesi}, M. and {Braun}, I. and {Briant}, T. and {Brillet}, A. and {Brinkmann}, M. and {Brockill}, P. and {Brockmueller}, E. and {Brooks}, A.~F. and {Brown}, B.~C. and {Brown}, D.~D. and {Brozzetti}, M.~L. and {Brunett}, S. and {Bruno}, G. and {Bruntz}, R. and {Bryant}, J. and {Bu}, Y. and {Bucci}, F. and {Buchanan}, J.},
        title = "{GWTC-4.0: Population Properties of Merging Compact Binaries}",
      journal = {\apjl},
         year = 2026,
        month = jul,
       volume = {1005},
       number = {2},
          eid = {L51},
        pages = {L51},
          doi = {10.3847/2041-8213/ae771e},
archivePrefix = {arXiv},
       eprint = {2508.18083},
 primaryClass = {astro-ph.HE},
       adsurl = {https://ui.adsabs.harvard.edu/abs/2026ApJ..1005L..51A}
}

@ARTICLE{Mapelli2017,
       author = {{Mapelli}, Michela and {Giacobbo}, Nicola and {Ripamonti}, Emanuele and {Spera}, Mario},
        title = "{The cosmic merger rate of stellar black hole binaries from the Illustris simulation}",
      journal = {\mnras},
         year = 2017,
        month = dec,
       volume = {472},
       number = {2},
        pages = {2422-2435},
          doi = {10.1093/mnras/stx2123},
archivePrefix = {arXiv},
       eprint = {1708.05722},
 primaryClass = {astro-ph.GA},
       adsurl = {https://ui.adsabs.harvard.edu/abs/2017MNRAS.472.2422M}
}

@ARTICLE{OShaughnessy2017,
       author = {{O'Shaughnessy}, R. and {Bellovary}, J.~M. and {Brooks}, A. and {Shen}, S. and {Governato}, F. and {Christensen}, C.~R.},
        title = "{The effects of host galaxy properties on merging compact binaries detectable by LIGO}",
      journal = {\mnras},
         year = 2017,
        month = jan,
       volume = {464},
       number = {3},
        pages = {2831-2839},
          doi = {10.1093/mnras/stw2550},
archivePrefix = {arXiv},
       eprint = {1609.06715},
 primaryClass = {astro-ph.GA},
       adsurl = {https://ui.adsabs.harvard.edu/abs/2017MNRAS.464.2831O}
}

@ARTICLE{Lamberts2016,
       author = {{Lamberts}, A. and {Garrison-Kimmel}, S. and {Clausen}, D.~R. and {Hopkins}, P.~F.},
        title = "{When and where did GW150914 form?}",
      journal = {\mnras},
         year = 2016,
        month = nov,
       volume = {463},
       number = {1},
        pages = {L31-L35},
          doi = {10.1093/mnrasl/slw152},
archivePrefix = {arXiv},
       eprint = {1605.08783},
 primaryClass = {astro-ph.HE},
       adsurl = {https://ui.adsabs.harvard.edu/abs/2016MNRAS.463L..31L}
}

@ARTICLE{Lamberts2018,
       author = {{Lamberts}, A. and {Garrison-Kimmel}, S. and {Hopkins}, P.~F. and {Quataert}, E. and {Bullock}, J.~S. and {Faucher-Gigu{\`e}re}, C.-A. and {Wetzel}, A. and {Kere{\v{s}}}, D. and {Drango}, K. and {Sanderson}, R.~E.},
        title = "{Predicting the binary black hole population of the Milky Way with cosmological simulations}",
      journal = {\mnras},
         year = 2018,
        month = oct,
       volume = {480},
       number = {2},
        pages = {2704-2718},
          doi = {10.1093/mnras/sty2035},
archivePrefix = {arXiv},
       eprint = {1801.03099},
 primaryClass = {astro-ph.GA},
       adsurl = {https://ui.adsabs.harvard.edu/abs/2018MNRAS.480.2704L}
}

@ARTICLE{Mapelli2018,
       author = {{Mapelli}, Michela and {Giacobbo}, Nicola},
        title = "{The cosmic merger rate of neutron stars and black holes}",
      journal = {\mnras},
         year = 2018,
        month = oct,
       volume = {479},
       number = {4},
        pages = {4391-4398},
          doi = {10.1093/mnras/sty1613},
archivePrefix = {arXiv},
       eprint = {1806.04866},
 primaryClass = {astro-ph.HE},
       adsurl = {https://ui.adsabs.harvard.edu/abs/2018MNRAS.479.4391M}
}

@ARTICLE{Artale2019,
       author = {{Artale}, M. Celeste and {Mapelli}, Michela and {Giacobbo}, Nicola and {Sabha}, Nadeen B. and {Spera}, Mario and {Santoliquido}, Filippo and {Bressan}, Alessandro},
        title = "{Host galaxies of merging compact objects: mass, star formation rate, metallicity, and colours}",
      journal = {\mnras},
         year = 2019,
        month = aug,
       volume = {487},
       number = {2},
        pages = {1675-1688},
          doi = {10.1093/mnras/stz1382},
archivePrefix = {arXiv},
       eprint = {1903.00083},
 primaryClass = {astro-ph.GA},
       adsurl = {https://ui.adsabs.harvard.edu/abs/2019MNRAS.487.1675A}
}

@ARTICLE{Belczynski2016,
       author = {{Belczynski}, Krzysztof and {Repetto}, Serena and {Holz}, Daniel E. and {O'Shaughnessy}, Richard and {Bulik}, Tomasz and {Berti}, Emanuele and {Fryer}, Christopher and {Dominik}, Michal},
        title = "{Compact Binary Merger Rates: Comparison with LIGO/Virgo Upper Limits}",
      journal = {\apj},
         year = 2016,
        month = mar,
       volume = {819},
       number = {2},
          eid = {108},
        pages = {108},
          doi = {10.3847/0004-637X/819/2/108},
archivePrefix = {arXiv},
       eprint = {1510.04615},
 primaryClass = {astro-ph.HE},
       adsurl = {https://ui.adsabs.harvard.edu/abs/2016ApJ...819..108B}
}

@ARTICLE{Cao2018,
       author = {{Cao}, Liang and {Lu}, Youjun and {Zhao}, Yuetong},
        title = "{Host galaxy properties of mergers of stellar binary black holes and their implications for advanced LIGO gravitational wave sources}",
      journal = {\mnras},
         year = 2018,
        month = mar,
       volume = {474},
       number = {4},
        pages = {4997-5007},
          doi = {10.1093/mnras/stx3087},
archivePrefix = {arXiv},
       eprint = {1711.09190},
 primaryClass = {astro-ph.GA},
       adsurl = {https://ui.adsabs.harvard.edu/abs/2018MNRAS.474.4997C}
}

@ARTICLE{Elbert2018,
       author = {{Elbert}, Oliver D. and {Bullock}, James S. and {Kaplinghat}, Manoj},
        title = "{Counting black holes: The cosmic stellar remnant population and implications for LIGO}",
      journal = {\mnras},
         year = 2018,
        month = jan,
       volume = {473},
       number = {1},
        pages = {1186-1194},
          doi = {10.1093/mnras/stx1959},
archivePrefix = {arXiv},
       eprint = {1703.02551},
 primaryClass = {astro-ph.GA},
       adsurl = {https://ui.adsabs.harvard.edu/abs/2018MNRAS.473.1186E}
}

@ARTICLE{Li2018,
       author = {{Li}, Shun-Sheng and {Mao}, Shude and {Zhao}, Yuetong and {Lu}, Youjun},
        title = "{Gravitational lensing of gravitational waves: a statistical perspective}",
      journal = {\mnras},
         year = 2018,
        month = may,
       volume = {476},
       number = {2},
        pages = {2220-2229},
          doi = {10.1093/mnras/sty411},
archivePrefix = {arXiv},
       eprint = {1802.05089},
 primaryClass = {astro-ph.CO},
       adsurl = {https://ui.adsabs.harvard.edu/abs/2018MNRAS.476.2220L}
}

@ARTICLE{Boco2019,
       author = {{Boco}, L. and {Lapi}, A. and {Goswami}, S. and {Perrotta}, F. and {Baccigalupi}, C. and {Danese}, L.},
        title = "{Merging Rates of Compact Binaries in Galaxies: Perspectives for Gravitational Wave Detections}",
      journal = {\apj},
         year = 2019,
        month = aug,
       volume = {881},
       number = {2},
          eid = {157},
        pages = {157},
          doi = {10.3847/1538-4357/ab328e},
archivePrefix = {arXiv},
       eprint = {1907.06841},
 primaryClass = {astro-ph.GA},
       adsurl = {https://ui.adsabs.harvard.edu/abs/2019ApJ...881..157B}
}

@ARTICLE{Boco2021,
       author = {{Boco}, L. and {Lapi}, A. and {Chruslinska}, M. and {Donevski}, D. and {Sicilia}, A. and {Danese}, L.},
        title = "{Evolution of Galaxy Star Formation and Metallicity: Impact on Double Compact Object Mergers}",
      journal = {\apj},
         year = 2021,
        month = feb,
       volume = {907},
       number = {2},
          eid = {110},
        pages = {110},
          doi = {10.3847/1538-4357/abd3a0},
archivePrefix = {arXiv},
       eprint = {2012.02800},
 primaryClass = {astro-ph.GA},
       adsurl = {https://ui.adsabs.harvard.edu/abs/2021ApJ...907..110B}
}

@ARTICLE{Chruslinska2021,
       author = {{Chru{\'s}li{\'n}ska}, Martyna and {Nelemans}, Gijs and {Boco}, Lumen and {Lapi}, Andrea},
        title = "{The impact of the FMR and starburst galaxies on the (low metallicity) cosmic star formation history}",
      journal = {\mnras},
         year = 2021,
        month = dec,
       volume = {508},
       number = {4},
        pages = {4994-5027},
          doi = {10.1093/mnras/stab2690},
archivePrefix = {arXiv},
       eprint = {2109.06187},
 primaryClass = {astro-ph.GA},
       adsurl = {https://ui.adsabs.harvard.edu/abs/2021MNRAS.508.4994C}
}

@ARTICLE{Chruslinska2019,
       author = {{Chruslinska}, Martyna and {Nelemans}, Gijs},
        title = "{Metallicity of stars formed throughout the cosmic history based on the observational properties of star-forming galaxies}",
      journal = {\mnras},
         year = 2019,
        month = oct,
       volume = {488},
       number = {4},
        pages = {5300-5326},
          doi = {10.1093/mnras/stz2057},
archivePrefix = {arXiv},
       eprint = {1907.11243},
 primaryClass = {astro-ph.GA},
       adsurl = {https://ui.adsabs.harvard.edu/abs/2019MNRAS.488.5300C}
}

@ARTICLE{Chruslinska2024,
       author = {{Chru{\'s}li{\'n}ska}, M. and {Pakmor}, R. and {Matthee}, J. and {Matsuno}, T.},
        title = "{Trading oxygen for iron. I. The [O/Fe]-specific star formation rate relation of galaxies}",
      journal = {\aap},
         year = 2024,
        month = jun,
       volume = {686},
          eid = {A186},
        pages = {A186},
          doi = {10.1051/0004-6361/202347602},
archivePrefix = {arXiv},
       eprint = {2308.00023},
 primaryClass = {astro-ph.GA},
       adsurl = {https://ui.adsabs.harvard.edu/abs/2024A&A...686A.186C}
}

@ARTICLE{Chruslinska2025,
       author = {{Chru{\'s}li{\'n}ska}, Martyna and {Curti}, Mirko and {Pakmor}, Ruediger and {De Cia}, Annalisa and {Matthee}, Jorryt and {Bhagwat}, Aniket and {Monty}, Stephanie},
        title = "{Trading oxygen for iron II. Oxygen- versus iron-dependent cosmic star formation history}",
      journal = {arXiv e-prints},
         year = 2025,
        month = nov,
          eid = {arXiv:2511.15782},
        pages = {arXiv:2511.15782},
          doi = {10.48550/arXiv.2511.15782},
archivePrefix = {arXiv},
       eprint = {2511.15782},
 primaryClass = {astro-ph.GA},
       adsurl = {https://ui.adsabs.harvard.edu/abs/2025arXiv251115782C}
}

@ARTICLE{Neijssel2019,
       author = {{Neijssel}, Coenraad J. and {Vigna-G{\'o}mez}, Alejandro and {Stevenson}, Simon and {Barrett}, Jim W. and {Gaebel}, Sebastian M. and {Broekgaarden}, Floor S. and {de Mink}, Selma E. and {Sz{\'e}csi}, Dorottya and {Vinciguerra}, Serena and {Mandel}, Ilya},
        title = "{The effect of the metallicity-specific star formation history on double compact object mergers}",
      journal = {\mnras},
         year = 2019,
        month = dec,
       volume = {490},
       number = {3},
        pages = {3740-3759},
          doi = {10.1093/mnras/stz2840},
archivePrefix = {arXiv},
       eprint = {1906.08136},
 primaryClass = {astro-ph.SR},
       adsurl = {https://ui.adsabs.harvard.edu/abs/2019MNRAS.490.3740N}
}

@ARTICLE{Santoliquido2020,
       author = {{Santoliquido}, Filippo and {Mapelli}, Michela and {Bouffanais}, Yann and {Giacobbo}, Nicola and {Di Carlo}, Ugo N. and {Rastello}, Sara and {Artale}, M. Celeste and {Ballone}, Alessandro},
        title = "{The Cosmic Merger Rate Density Evolution of Compact Binaries Formed in Young Star Clusters and in Isolated Binaries}",
      journal = {\apj},
         year = 2020,
        month = aug,
       volume = {898},
       number = {2},
          eid = {152},
        pages = {152},
          doi = {10.3847/1538-4357/ab9b78},
archivePrefix = {arXiv},
       eprint = {2004.09533},
 primaryClass = {astro-ph.HE},
       adsurl = {https://ui.adsabs.harvard.edu/abs/2020ApJ...898..152S}
}

@ARTICLE{Santoliquido2022,
       author = {{Santoliquido}, Filippo and {Mapelli}, Michela and {Artale}, M. Celeste and {Boco}, Lumen},
        title = "{Modelling the host galaxies of binary compact object mergers with observational scaling relations}",
      journal = {\mnras},
         year = 2022,
        month = nov,
       volume = {516},
       number = {3},
        pages = {3297-3317},
          doi = {10.1093/mnras/stac2384},
archivePrefix = {arXiv},
       eprint = {2205.05099},
 primaryClass = {astro-ph.HE},
       adsurl = {https://ui.adsabs.harvard.edu/abs/2022MNRAS.516.3297S}
}

@ARTICLE{Broekgaarden2022,
       author = {{Broekgaarden}, Floor S. and {Berger}, Edo and {Stevenson}, Simon and {Justham}, Stephen and {Mandel}, Ilya and {Chru{\'s}li{\'n}ska}, Martyna and {van Son}, Lieke A.~C. and {Wagg}, Tom and {Vigna-G{\'o}mez}, Alejandro and {de Mink}, Selma E. and {Chattopadhyay}, Debatri and {Neijssel}, Coenraad J.},
        title = "{Impact of massive binary star and cosmic evolution on gravitational wave observations - II. Double compact object rates and properties}",
      journal = {\mnras},
         year = 2022,
        month = nov,
       volume = {516},
       number = {4},
        pages = {5737-5761},
          doi = {10.1093/mnras/stac1677},
archivePrefix = {arXiv},
       eprint = {2112.05763},
 primaryClass = {astro-ph.HE},
       adsurl = {https://ui.adsabs.harvard.edu/abs/2022MNRAS.516.5737B}
}

@ARTICLE{Sgalletta2025,
       author = {{Sgalletta}, Cecilia and {Mapelli}, Michela and {Boco}, Lumen and {Santoliquido}, Filippo and {Artale}, M. Celeste and {Iorio}, Giuliano and {Lapi}, Andrea and {Spera}, Mario},
        title = "{The more accurately the metal-dependent star formation rate is modeled, the larger the predicted excess of binary black hole mergers}",
      journal = {\aap},
         year = 2025,
        month = jun,
       volume = {698},
          eid = {A144},
        pages = {A144},
          doi = {10.1051/0004-6361/202452757},
archivePrefix = {arXiv},
       eprint = {2410.21401},
 primaryClass = {astro-ph.HE},
       adsurl = {https://ui.adsabs.harvard.edu/abs/2025A&A...698A.144S}
}

@ARTICLE{vanson2023,
       author = {{van Son}, L.~A.~C. and {de Mink}, S.~E. and {Chru{\'s}li{\'n}ska}, M. and {Conroy}, C. and {Pakmor}, R. and {Hernquist}, L.},
        title = "{The Locations of Features in the Mass Distribution of Merging Binary Black Holes Are Robust against Uncertainties in the Metallicity-dependent Cosmic Star Formation History}",
      journal = {\apj},
         year = 2023,
        month = may,
       volume = {948},
       number = {2},
          eid = {105},
        pages = {105},
          doi = {10.3847/1538-4357/acbf51},
archivePrefix = {arXiv},
       eprint = {2209.03385},
 primaryClass = {astro-ph.GA},
       adsurl = {https://ui.adsabs.harvard.edu/abs/2023ApJ...948..105V}
}

@ARTICLE{Andrews2013,
       author = {{Andrews}, Brett H. and {Martini}, Paul},
        title = "{The Mass-Metallicity Relation with the Direct Method on Stacked Spectra of SDSS Galaxies}",
      journal = {\apj},
         year = 2013,
        month = mar,
       volume = {765},
       number = {2},
          eid = {140},
        pages = {140},
          doi = {10.1088/0004-637X/765/2/140},
archivePrefix = {arXiv},
       eprint = {1211.3418},
 primaryClass = {astro-ph.CO},
       adsurl = {https://ui.adsabs.harvard.edu/abs/2013ApJ...765..140A}
}

@ARTICLE{Grevesse1998,
       author = {{Grevesse}, N. and {Sauval}, A.~J.},
        title = "{Standard Solar Composition}",
      journal = {\ssr},
         year = 1998,
        month = may,
       volume = {85},
        pages = {161-174},
          doi = {10.1023/A:1005161325181},
       adsurl = {https://ui.adsabs.harvard.edu/abs/1998SSRv...85..161G}
}

@ARTICLE{Steidel2014,
       author = {{Steidel}, Charles C. and {Rudie}, Gwen C. and {Strom}, Allison L. and {Pettini}, Max and {Reddy}, Naveen A. and {Shapley}, Alice E. and {Trainor}, Ryan F. and {Erb}, Dawn K. and {Turner}, Monica L. and {Konidaris}, Nicholas P. and {Kulas}, Kristin R. and {Mace}, Gregory and {Matthews}, Keith and {McLean}, Ian S.},
        title = "{Strong Nebular Line Ratios in the Spectra of z \raisebox{-0.5ex}\textasciitilde 2-3 Star Forming Galaxies: First Results from KBSS-MOSFIRE}",
      journal = {\apj},
         year = 2014,
        month = nov,
       volume = {795},
       number = {2},
          eid = {165},
        pages = {165},
          doi = {10.1088/0004-637X/795/2/165},
archivePrefix = {arXiv},
       eprint = {1405.5473},
 primaryClass = {astro-ph.GA},
       adsurl = {https://ui.adsabs.harvard.edu/abs/2014ApJ...795..165S}
}

@ARTICLE{Steidel2016,
       author = {{Steidel}, Charles C. and {Strom}, Allison L. and {Pettini}, Max and {Rudie}, Gwen C. and {Reddy}, Naveen A. and {Trainor}, Ryan F.},
        title = "{Reconciling the Stellar and Nebular Spectra of High-redshift Galaxies}",
      journal = {\apj},
         year = 2016,
        month = aug,
       volume = {826},
       number = {2},
          eid = {159},
        pages = {159},
          doi = {10.3847/0004-637X/826/2/159},
archivePrefix = {arXiv},
       eprint = {1605.07186},
 primaryClass = {astro-ph.GA},
       adsurl = {https://ui.adsabs.harvard.edu/abs/2016ApJ...826..159S}
}

@ARTICLE{Topping2020,
       author = {{Topping}, Michael W. and {Shapley}, Alice E. and {Reddy}, Naveen A. and {Sanders}, Ryan L. and {Coil}, Alison L. and {Kriek}, Mariska and {Mobasher}, Bahram and {Siana}, Brian},
        title = "{The MOSDEF-LRIS Survey: The connection between massive stars and ionized gas in individual galaxies at z {\ensuremath{\sim}} 2}",
      journal = {\mnras},
         year = 2020,
        month = dec,
       volume = {499},
       number = {2},
        pages = {1652-1665},
          doi = {10.1093/mnras/staa2941},
archivePrefix = {arXiv},
       eprint = {2008.02282},
 primaryClass = {astro-ph.GA},
       adsurl = {https://ui.adsabs.harvard.edu/abs/2020MNRAS.499.1652T}
}

@ARTICLE{Sanders2021,
       author = {{Sanders}, Ryan L. and {Shapley}, Alice E. and {Jones}, Tucker and {Reddy}, Naveen A. and {Kriek}, Mariska and {Siana}, Brian and {Coil}, Alison L. and {Mobasher}, Bahram and {Shivaei}, Irene and {Dav{\'e}}, Romeel and {Azadi}, Mojegan and {Price}, Sedona H. and {Leung}, Gene and {Freeman}, William R. and {Fetherolf}, Tara and {de Groot}, Laura and {Zick}, Tom and {Barro}, Guillermo},
        title = "{The MOSDEF Survey: The Evolution of the Mass-Metallicity Relation from z = 0 to z 3.3}",
      journal = {\apj},
         year = 2021,
        month = jun,
       volume = {914},
       number = {1},
          eid = {19},
        pages = {19},
          doi = {10.3847/1538-4357/abf4c1},
archivePrefix = {arXiv},
       eprint = {2009.07292},
 primaryClass = {astro-ph.GA},
       adsurl = {https://ui.adsabs.harvard.edu/abs/2021ApJ...914...19S}
}

@ARTICLE{Cullen2021,
       author = {{Cullen}, F. and {Shapley}, A.~E. and {McLure}, R.~J. and {Dunlop}, J.~S. and {Sanders}, R.~L. and {Topping}, M.~W. and {Reddy}, N.~A. and {Amor{\'\i}n}, R. and {Begley}, R. and {Bolzonella}, M. and {Calabr{\`o}}, A. and {Carnall}, A.~C. and {Castellano}, M. and {Cimatti}, A. and {Cirasuolo}, M. and {Cresci}, G. and {Fontana}, A. and {Fontanot}, F. and {Garilli}, B. and {Guaita}, L. and {Hamadouche}, M. and {Hathi}, N.~P. and {Mannucci}, F. and {McLeod}, D.~J. and {Pentericci}, L. and {Saxena}, A. and {Talia}, M. and {Zamorani}, G.},
        title = "{The NIRVANDELS Survey: a robust detection of {\ensuremath{\alpha}}-enhancement in star-forming galaxies at z ≃ 3.4}",
      journal = {\mnras},
         year = 2021,
        month = jul,
       volume = {505},
       number = {1},
        pages = {903-920},
          doi = {10.1093/mnras/stab1340},
archivePrefix = {arXiv},
       eprint = {2103.06300},
 primaryClass = {astro-ph.GA},
       adsurl = {https://ui.adsabs.harvard.edu/abs/2021MNRAS.505..903C}
}

@ARTICLE{Strom2022,
       author = {{Strom}, Allison L. and {Rudie}, Gwen C. and {Steidel}, Charles C. and {Trainor}, Ryan F.},
        title = "{Chemical Abundance Scaling Relations for Multiple Elements in z ≃ 2-3 Star-forming Galaxies}",
      journal = {\apj},
         year = 2022,
        month = feb,
       volume = {925},
       number = {2},
          eid = {116},
        pages = {116},
          doi = {10.3847/1538-4357/ac38a3},
archivePrefix = {arXiv},
       eprint = {2111.06410},
 primaryClass = {astro-ph.GA},
       adsurl = {https://ui.adsabs.harvard.edu/abs/2022ApJ...925..116S}
}

@ARTICLE{Thomas2005,
       author = {{Thomas}, Daniel and {Maraston}, Claudia and {Bender}, Ralf and {Mendes de Oliveira}, Claudia},
        title = "{The Epochs of Early-Type Galaxy Formation as a Function of Environment}",
      journal = {\apj},
         year = 2005,
        month = mar,
       volume = {621},
       number = {2},
        pages = {673-694},
          doi = {10.1086/426932},
archivePrefix = {arXiv},
       eprint = {astro-ph/0410209},
 primaryClass = {astro-ph},
       adsurl = {https://ui.adsabs.harvard.edu/abs/2005ApJ...621..673T}
}

@ARTICLE{Thomas2010,
       author = {{Thomas}, Daniel and {Maraston}, Claudia and {Schawinski}, Kevin and {Sarzi}, Marc and {Silk}, Joseph},
        title = "{Environment and self-regulation in galaxy formation}",
      journal = {\mnras},
         year = 2010,
        month = jun,
       volume = {404},
       number = {4},
        pages = {1775-1789},
          doi = {10.1111/j.1365-2966.2010.16427.x},
archivePrefix = {arXiv},
       eprint = {0912.0259},
 primaryClass = {astro-ph.CO},
       adsurl = {https://ui.adsabs.harvard.edu/abs/2010MNRAS.404.1775T}
}

@ARTICLE{Johansson2012,
       author = {{Johansson}, Jonas and {Thomas}, Daniel and {Maraston}, Claudia},
        title = "{Chemical element ratios of Sloan Digital Sky Survey early-type galaxies}",
      journal = {\mnras},
         year = 2012,
        month = apr,
       volume = {421},
       number = {3},
        pages = {1908-1926},
          doi = {10.1111/j.1365-2966.2011.20316.x},
archivePrefix = {arXiv},
       eprint = {1112.0322},
 primaryClass = {astro-ph.CO},
       adsurl = {https://ui.adsabs.harvard.edu/abs/2012MNRAS.421.1908J}
}

@ARTICLE{Alonso-Alvarez2025,
       author = {{{\'A}lvarez}, Carlos A. and {Cueli}, Marcos M. and {Bressan}, Alessandro and {Boco}, Lumen and {Haridasu}, Balakrishna S. and {Bosi}, Michele and {Danese}, Luigi and {Lapi}, Andrea},
        title = "{Cosmography via stellar archaeology of low-redshift early-type galaxies from SDSS}",
      journal = {\aap},
         year = 2025,
        month = nov,
       volume = {703},
          eid = {A26},
        pages = {A26},
          doi = {10.1051/0004-6361/202555727},
archivePrefix = {arXiv},
       eprint = {2509.04224},
 primaryClass = {astro-ph.CO},
       adsurl = {https://ui.adsabs.harvard.edu/abs/2025A&A...703A..26A}
}

@ARTICLE{Sana2012,
       author = {{Sana}, H. and {de Mink}, S.~E. and {de Koter}, A. and {Langer}, N. and {Evans}, C.~J. and {Gieles}, M. and {Gosset}, E. and {Izzard}, R.~G. and {Le Bouquin}, J.-B. and {Schneider}, F.~R.~N.},
        title = "{Binary Interaction Dominates the Evolution of Massive Stars}",
      journal = {Science},
         year = 2012,
        month = jul,
       volume = {337},
       number = {6093},
        pages = {444},
          doi = {10.1126/science.1223344},
archivePrefix = {arXiv},
       eprint = {1207.6397},
 primaryClass = {astro-ph.SR},
       adsurl = {https://ui.adsabs.harvard.edu/abs/2012Sci...337..444S}
}

@ARTICLE{Ilbert2013,
       author = {{Ilbert}, O. and {McCracken}, H.~J. and {Le F{\`e}vre}, O. and {Capak}, P. and {Dunlop}, J. and {Karim}, A. and {Renzini}, M.~A. and {Caputi}, K. and {Boissier}, S. and {Arnouts}, S. and {Aussel}, H. and {Comparat}, J. and {Guo}, Q. and {Hudelot}, P. and {Kartaltepe}, J. and {Kneib}, J.~P. and {Krogager}, J.~K. and {Le Floc'h}, E. and {Lilly}, S. and {Mellier}, Y. and {Milvang-Jensen}, B. and {Moutard}, T. and {Onodera}, M. and {Richard}, J. and {Salvato}, M. and {Sanders}, D.~B. and {Scoville}, N. and {Silverman}, J.~D. and {Taniguchi}, Y. and {Tasca}, L. and {Thomas}, R. and {Toft}, S. and {Tresse}, L. and {Vergani}, D. and {Wolk}, M. and {Zirm}, A.},
        title = "{Mass assembly in quiescent and star-forming galaxies since z ≃ 4 from UltraVISTA}",
      journal = {\aap},
         year = 2013,
        month = aug,
       volume = {556},
          eid = {A55},
        pages = {A55},
          doi = {10.1051/0004-6361/201321100},
archivePrefix = {arXiv},
       eprint = {1301.3157},
 primaryClass = {astro-ph.CO},
       adsurl = {https://ui.adsabs.harvard.edu/abs/2013A&A...556A..55I}
}

@ARTICLE{Muzzin2013,
       author = {{Muzzin}, Adam and {Marchesini}, Danilo and {Stefanon}, Mauro and {Franx}, Marijn and {McCracken}, Henry J. and {Milvang-Jensen}, Bo and {Dunlop}, James S. and {Fynbo}, J.~P.~U. and {Brammer}, Gabriel and {Labb{\'e}}, Ivo and {van Dokkum}, Pieter G.},
        title = "{The Evolution of the Stellar Mass Functions of Star-forming and Quiescent Galaxies to z = 4 from the COSMOS/UltraVISTA Survey}",
      journal = {\apj},
         year = 2013,
        month = nov,
       volume = {777},
       number = {1},
          eid = {18},
        pages = {18},
          doi = {10.1088/0004-637X/777/1/18},
archivePrefix = {arXiv},
       eprint = {1303.4409},
 primaryClass = {astro-ph.CO},
       adsurl = {https://ui.adsabs.harvard.edu/abs/2013ApJ...777...18M}
}

@ARTICLE{Tomczak2014,
       author = {{Tomczak}, Adam R. and {Quadri}, Ryan F. and {Tran}, Kim-Vy H. and {Labb{\'e}}, Ivo and {Straatman}, Caroline M.~S. and {Papovich}, Casey and {Glazebrook}, Karl and {Allen}, Rebecca and {Brammer}, Gabriel B. and {Kacprzak}, Glenn G. and {Kawinwanichakij}, Lalitwadee and {Kelson}, Daniel D. and {McCarthy}, Patrick J. and {Mehrtens}, Nicola and {Monson}, Andrew J. and {Persson}, S. Eric and {Spitler}, Lee R. and {Tilvi}, Vithal and {van Dokkum}, Pieter},
        title = "{Galaxy Stellar Mass Functions from ZFOURGE/CANDELS: An Excess of Low-mass Galaxies since z = 2 and the Rapid Buildup of Quiescent Galaxies}",
      journal = {\apj},
         year = 2014,
        month = mar,
       volume = {783},
       number = {2},
          eid = {85},
        pages = {85},
          doi = {10.1088/0004-637X/783/2/85},
archivePrefix = {arXiv},
       eprint = {1309.5972},
 primaryClass = {astro-ph.CO},
       adsurl = {https://ui.adsabs.harvard.edu/abs/2014ApJ...783...85T}
}

@ARTICLE{Davidzon2017,
       author = {{Davidzon}, I. and {Ilbert}, O. and {Laigle}, C. and {Coupon}, J. and {McCracken}, H.~J. and {Delvecchio}, I. and {Masters}, D. and {Capak}, P. and {Hsieh}, B.~C. and {Le F{\`e}vre}, O. and {Tresse}, L. and {Bethermin}, M. and {Chang}, Y.-Y. and {Faisst}, A.~L. and {Le Floc'h}, E. and {Steinhardt}, C. and {Toft}, S. and {Aussel}, H. and {Dubois}, C. and {Hasinger}, G. and {Salvato}, M. and {Sanders}, D.~B. and {Scoville}, N. and {Silverman}, J.~D.},
        title = "{The COSMOS2015 galaxy stellar mass function . Thirteen billion years of stellar mass assembly in ten snapshots}",
      journal = {\aap},
         year = 2017,
        month = sep,
       volume = {605},
          eid = {A70},
        pages = {A70},
          doi = {10.1051/0004-6361/201730419},
archivePrefix = {arXiv},
       eprint = {1701.02734},
 primaryClass = {astro-ph.GA},
       adsurl = {https://ui.adsabs.harvard.edu/abs/2017A&A...605A..70D}
}

@ARTICLE{Weaver2023,
       author = {{Weaver}, J.~R. and {Davidzon}, I. and {Toft}, S. and {Ilbert}, O. and {McCracken}, H.~J. and {Gould}, K.~M.~L. and {Jespersen}, C.~K. and {Steinhardt}, C. and {Lagos}, C.~D.~P. and {Capak}, P.~L. and {Casey}, C.~M. and {Chartab}, N. and {Faisst}, A.~L. and {Hayward}, C.~C. and {Kartaltepe}, J.~S. and {Kauffmann}, O.~B. and {Koekemoer}, A.~M. and {Kokorev}, V. and {Laigle}, C. and {Liu}, D. and {Long}, A. and {Magdis}, G.~E. and {McPartland}, C.~J.~R. and {Milvang-Jensen}, B. and {Mobasher}, B. and {Moneti}, A. and {Peng}, Y. and {Sanders}, D.~B. and {Shuntov}, M. and {Sneppen}, A. and {Valentino}, F. and {Zalesky}, L. and {Zamorani}, G.},
        title = "{COSMOS2020: The galaxy stellar mass function. The assembly and star formation cessation of galaxies at 0.2< z {\ensuremath{\leq}} 7.5}",
      journal = {\aap},
         year = 2023,
        month = sep,
       volume = {677},
          eid = {A184},
        pages = {A184},
          doi = {10.1051/0004-6361/202245581},
archivePrefix = {arXiv},
       eprint = {2212.02512},
 primaryClass = {astro-ph.GA},
       adsurl = {https://ui.adsabs.harvard.edu/abs/2023A&A...677A.184W}
}

@ARTICLE{Shuntov2025,
       author = {{Shuntov}, Marko and {Akins}, Hollis B. and {Paquereau}, Louise and {Casey}, Caitlin M. and {Ilbert}, Olivier and {Arango-Toro}, Rafael C. and {McCracken}, Henry Joy and {Franco}, Maximilien and {Harish}, Santosh and {Kartaltepe}, Jeyhan S. and {Koekemoer}, Anton M. and {Yang}, Lilan and {Huertas-Company}, Marc and {Berman}, Edward M. and {McCleary}, Jacqueline E. and {Toft}, Sune and {Gavazzi}, Rapha{\"e}l and {Achenbach}, Mark J. and {Bertin}, Emmanuel and {Brinch}, Malte and {Champagne}, Jackie and {Chartab}, Nima and {Drakos}, Nicole E. and {Egami}, Eiichi and {Endsley}, Ryan and {Faisst}, Andreas L. and {Fan}, Xiaohui and {Flayhart}, Carter and {Hartley}, William G. and {Hatamnia}, Hossein and {Gozaliasl}, Ghassem and {Gentile}, Fabrizio and {Jermann}, Iris and {Jin}, Shuowen and {Kakiichi}, Koki and {Khostovan}, Ali Ahmad and {K{\"u}mmel}, Martin and {Laigle}, Clotilde and {Laishram}, Ronaldo and {Lambrides}, Erini and {Liu}, Daizhong and {Lyu}, Jianwei and {Magdis}, Georgios and {Mobasher}, Bahram and {Moutard}, Thibaud and {Renzini}, Alvio and {Rich}, R. Michael and {Sanders}, David B. and {Sattari}, Zahra and {Robertson}, Brant E. and {Schefer}, Marc and {Scognamiglio}, Diana and {Scoville}, Nick and {Silverman}, John D. and {Taamoli}, Sina and {Trakhtenbrot}, Benny and {Valentino}, Francesco and {Wang}, Feige and {Weaver}, John R. and {Yang}, Jinyi},
        title = "{COSMOS2025: The COSMOS-Web galaxy catalog of photometry, morphology, redshifts, and physical parameters from JWST, HST, and ground-based imaging}",
      journal = {\aap},
         year = 2025,
        month = dec,
       volume = {704},
          eid = {A339},
        pages = {A339},
          doi = {10.1051/0004-6361/202555799},
archivePrefix = {arXiv},
       eprint = {2506.03243},
 primaryClass = {astro-ph.GA},
       adsurl = {https://ui.adsabs.harvard.edu/abs/2025A&A...704A.339S}
}

@ARTICLE{Daddi2007,
       author = {{Daddi}, E. and {Alexander}, D.~M. and {Dickinson}, M. and {Gilli}, R. and {Renzini}, A. and {Elbaz}, D. and {Cimatti}, A. and {Chary}, R. and {Frayer}, D. and {Bauer}, F.~E. and {Brandt}, W.~N. and {Giavalisco}, M. and {Grogin}, N.~A. and {Huynh}, M. and {Kurk}, J. and {Mignoli}, M. and {Morrison}, G. and {Pope}, A. and {Ravindranath}, S.},
        title = "{Multiwavelength Study of Massive Galaxies at z\raisebox{-0.5ex}\textasciitilde2. II. Widespread Compton-thick Active Galactic Nuclei and the Concurrent Growth of Black Holes and Bulges}",
      journal = {\apj},
         year = 2007,
        month = nov,
       volume = {670},
       number = {1},
        pages = {173-189},
          doi = {10.1086/521820},
archivePrefix = {arXiv},
       eprint = {0705.2832},
 primaryClass = {astro-ph},
       adsurl = {https://ui.adsabs.harvard.edu/abs/2007ApJ...670..173D}
}

@ARTICLE{Speagle2014,
       author = {{Speagle}, J.~S. and {Steinhardt}, C.~L. and {Capak}, P.~L. and {Silverman}, J.~D.},
        title = "{A Highly Consistent Framework for the Evolution of the Star-Forming ``Main Sequence'' from z \raisebox{-0.5ex}\textasciitilde 0-6}",
      journal = {\apjs},
         year = 2014,
        month = oct,
       volume = {214},
       number = {2},
          eid = {15},
        pages = {15},
          doi = {10.1088/0067-0049/214/2/15},
archivePrefix = {arXiv},
       eprint = {1405.2041},
 primaryClass = {astro-ph.GA},
       adsurl = {https://ui.adsabs.harvard.edu/abs/2014ApJS..214...15S}
}

@ARTICLE{Popesso2023,
       author = {{Popesso}, P. and {Concas}, A. and {Cresci}, G. and {Belli}, S. and {Rodighiero}, G. and {Inami}, H. and {Dickinson}, M. and {Ilbert}, O. and {Pannella}, M. and {Elbaz}, D.},
        title = "{The main sequence of star-forming galaxies across cosmic times}",
      journal = {\mnras},
         year = 2023,
        month = feb,
       volume = {519},
       number = {1},
        pages = {1526-1544},
          doi = {10.1093/mnras/stac3214},
archivePrefix = {arXiv},
       eprint = {2203.10487},
 primaryClass = {astro-ph.GA},
       adsurl = {https://ui.adsabs.harvard.edu/abs/2023MNRAS.519.1526P}
}

@ARTICLE{Whitaker2014,
       author = {{Whitaker}, Katherine E. and {Franx}, Marijn and {Leja}, Joel and {van Dokkum}, Pieter G. and {Henry}, Alaina and {Skelton}, Rosalind E. and {Fumagalli}, Mattia and {Momcheva}, Ivelina G. and {Brammer}, Gabriel B. and {Labb{\'e}}, Ivo and {Nelson}, Erica J. and {Rigby}, Jane R.},
        title = "{Constraining the Low-mass Slope of the Star Formation Sequence at 0.5 < z < 2.5}",
      journal = {\apj},
         year = 2014,
        month = nov,
       volume = {795},
       number = {2},
          eid = {104},
        pages = {104},
          doi = {10.1088/0004-637X/795/2/104},
archivePrefix = {arXiv},
       eprint = {1407.1843},
 primaryClass = {astro-ph.GA},
       adsurl = {https://ui.adsabs.harvard.edu/abs/2014ApJ...795..104W}
}

@ARTICLE{Pantoni2019,
       author = {{Pantoni}, L. and {Lapi}, A. and {Massardi}, M. and {Goswami}, S. and {Danese}, L.},
        title = "{New Analytic Solutions for Galaxy Evolution: Gas, Stars, Metals, and Dust in Local ETGs and Their High-z Star-forming Progenitors}",
      journal = {\apj},
         year = 2019,
        month = aug,
       volume = {880},
       number = {2},
          eid = {129},
        pages = {129},
          doi = {10.3847/1538-4357/ab2adc},
archivePrefix = {arXiv},
       eprint = {1906.07458},
 primaryClass = {astro-ph.GA},
       adsurl = {https://ui.adsabs.harvard.edu/abs/2019ApJ...880..129P}
}

@ARTICLE{Lapi2020,
       author = {{Lapi}, A. and {Pantoni}, L. and {Boco}, L. and {Danese}, L.},
        title = "{New Analytic Solutions for Galaxy Evolution. II. Wind Recycling, Galactic Fountains, and Late-type Galaxies}",
      journal = {\apj},
         year = 2020,
        month = jul,
       volume = {897},
       number = {1},
          eid = {81},
        pages = {81},
          doi = {10.3847/1538-4357/ab9812},
archivePrefix = {arXiv},
       eprint = {2006.01643},
 primaryClass = {astro-ph.GA},
       adsurl = {https://ui.adsabs.harvard.edu/abs/2020ApJ...897...81L}
}

@ARTICLE{Bosi2025,
       author = {{Bosi}, Michele and {Lapi}, Andrea and {Boco}, Lumen and {Alvarez}, Carlos A. and {Cueli}, Marcos M. and {Antinozzi}, Giovanni and {Behiri}, Meriem and {Giulietti}, Marika and {Massardi}, Marcella and {Spera}, Mario and {Bressan}, Alessandro and {Baccigalupi}, Carlo and {Danese}, Luigi},
        title = "{StAGE: Stellar Archaeology-driven Galaxy Evolution}",
      journal = {\apj},
         year = 2025,
        month = may,
       volume = {984},
       number = {2},
          eid = {117},
        pages = {117},
          doi = {10.3847/1538-4357/adc721},
archivePrefix = {arXiv},
       eprint = {2503.22543},
 primaryClass = {astro-ph.GA},
       adsurl = {https://ui.adsabs.harvard.edu/abs/2025ApJ...984..117B}
}

@ARTICLE{Rodighiero2011,
       author = {{Rodighiero}, G. and {Daddi}, E. and {Baronchelli}, I. and {Cimatti}, A. and {Renzini}, A. and {Aussel}, H. and {Popesso}, P. and {Lutz}, D. and {Andreani}, P. and {Berta}, S. and {Cava}, A. and {Elbaz}, D. and {Feltre}, A. and {Fontana}, A. and {F{\"o}rster Schreiber}, N.~M. and {Franceschini}, A. and {Genzel}, R. and {Grazian}, A. and {Gruppioni}, C. and {Ilbert}, O. and {Le Floch}, E. and {Magdis}, G. and {Magliocchetti}, M. and {Magnelli}, B. and {Maiolino}, R. and {McCracken}, H. and {Nordon}, R. and {Poglitsch}, A. and {Santini}, P. and {Pozzi}, F. and {Riguccini}, L. and {Tacconi}, L.~J. and {Wuyts}, S. and {Zamorani}, G.},
        title = "{The Lesser Role of Starbursts in Star Formation at z = 2}",
      journal = {\apjl},
         year = 2011,
        month = oct,
       volume = {739},
       number = {2},
          eid = {L40},
        pages = {L40},
          doi = {10.1088/2041-8205/739/2/L40},
archivePrefix = {arXiv},
       eprint = {1108.0933},
 primaryClass = {astro-ph.CO},
       adsurl = {https://ui.adsabs.harvard.edu/abs/2011ApJ...739L..40R}
}

@ARTICLE{Rodighiero2015,
       author = {{Rodighiero}, G. and {Brusa}, M. and {Daddi}, E. and {Negrello}, M. and {Mullaney}, J.~R. and {Delvecchio}, I. and {Lutz}, D. and {Renzini}, A. and {Franceschini}, A. and {Baronchelli}, I. and {Pozzi}, F. and {Gruppioni}, C. and {Strazzullo}, V. and {Cimatti}, A. and {Silverman}, J.},
        title = "{Relationship between Star Formation Rate and Black Hole Accretion At Z = 2: the Different Contributions in Quiescent, Normal, and Starburst Galaxies}",
      journal = {\apjl},
         year = 2015,
        month = feb,
       volume = {800},
       number = {1},
          eid = {L10},
        pages = {L10},
          doi = {10.1088/2041-8205/800/1/L10},
archivePrefix = {arXiv},
       eprint = {1501.04634},
 primaryClass = {astro-ph.GA},
       adsurl = {https://ui.adsabs.harvard.edu/abs/2015ApJ...800L..10R}
}

@ARTICLE{Schreiber2015,
       author = {{Schreiber}, C. and {Pannella}, M. and {Elbaz}, D. and {B{\'e}thermin}, M. and {Inami}, H. and {Dickinson}, M. and {Magnelli}, B. and {Wang}, T. and {Aussel}, H. and {Daddi}, E. and {Juneau}, S. and {Shu}, X. and {Sargent}, M.~T. and {Buat}, V. and {Faber}, S.~M. and {Ferguson}, H.~C. and {Giavalisco}, M. and {Koekemoer}, A.~M. and {Magdis}, G. and {Morrison}, G.~E. and {Papovich}, C. and {Santini}, P. and {Scott}, D.},
        title = "{The Herschel view of the dominant mode of galaxy growth from z = 4 to the present day}",
      journal = {\aap},
         year = 2015,
        month = mar,
       volume = {575},
          eid = {A74},
        pages = {A74},
          doi = {10.1051/0004-6361/201425017},
archivePrefix = {arXiv},
       eprint = {1409.5433},
 primaryClass = {astro-ph.GA},
       adsurl = {https://ui.adsabs.harvard.edu/abs/2015A&A...575A..74S}
}

@ARTICLE{Dunlop2017,
       author = {{Dunlop}, J.~S. and {McLure}, R.~J. and {Biggs}, A.~D. and {Geach}, J.~E. and {Micha{\l}owski}, M.~J. and {Ivison}, R.~J. and {Rujopakarn}, W. and {van Kampen}, E. and {Kirkpatrick}, A. and {Pope}, A. and {Scott}, D. and {Swinbank}, A.~M. and {Targett}, T.~A. and {Aretxaga}, I. and {Austermann}, J.~E. and {Best}, P.~N. and {Bruce}, V.~A. and {Chapin}, E.~L. and {Charlot}, S. and {Cirasuolo}, M. and {Coppin}, K. and {Ellis}, R.~S. and {Finkelstein}, S.~L. and {Hayward}, C.~C. and {Hughes}, D.~H. and {Ibar}, E. and {Jagannathan}, P. and {Khochfar}, S. and {Koprowski}, M.~P. and {Narayanan}, D. and {Nyland}, K. and {Papovich}, C. and {Peacock}, J.~A. and {Rieke}, G.~H. and {Robertson}, B. and {Vernstrom}, T. and {Werf}, P.~P. van der and {Wilson}, G.~W. and {Yun}, M.},
        title = "{A deep ALMA image of the Hubble Ultra Deep Field}",
      journal = {\mnras},
         year = 2017,
        month = apr,
       volume = {466},
       number = {1},
        pages = {861-883},
          doi = {10.1093/mnras/stw3088},
archivePrefix = {arXiv},
       eprint = {1606.00227},
 primaryClass = {astro-ph.GA},
       adsurl = {https://ui.adsabs.harvard.edu/abs/2017MNRAS.466..861D}
}

@ARTICLE{Bisigello2018,
       author = {{Bisigello}, L. and {Caputi}, K.~I. and {Grogin}, N. and {Koekemoer}, A.},
        title = "{Analysis of the SFR-M$^{{\ensuremath{*}}}$ plane at z < 3: single fitting versus multi-Gaussian decomposition}",
      journal = {\aap},
         year = 2018,
        month = jan,
       volume = {609},
          eid = {A82},
        pages = {A82},
          doi = {10.1051/0004-6361/201731399},
archivePrefix = {arXiv},
       eprint = {1706.06154},
 primaryClass = {astro-ph.GA},
       adsurl = {https://ui.adsabs.harvard.edu/abs/2018A&A...609A..82B}
}

@ARTICLE{Caputi2017,
       author = {{Caputi}, K.~I. and {Deshmukh}, S. and {Ashby}, M.~L.~N. and {Cowley}, W.~I. and {Bisigello}, L. and {Fazio}, G.~G. and {Fynbo}, J.~P.~U. and {Le F{\`e}vre}, O. and {Milvang-Jensen}, B. and {Ilbert}, O.},
        title = "{Star Formation in Galaxies at z {\ensuremath{\sim}} 4-5 from the SMUVS Survey: A Clear Starburst/Main-sequence Bimodality for H{\ensuremath{\alpha}} Emitters on the SFR-M* Plane}",
      journal = {\apj},
         year = 2017,
        month = nov,
       volume = {849},
       number = {1},
          eid = {45},
        pages = {45},
          doi = {10.3847/1538-4357/aa901e},
archivePrefix = {arXiv},
       eprint = {1705.06179},
 primaryClass = {astro-ph.GA},
       adsurl = {https://ui.adsabs.harvard.edu/abs/2017ApJ...849...45C}
}

@ARTICLE{Mancuso2016b,
       author = {{Mancuso}, C. and {Lapi}, A. and {Shi}, J. and {Cai}, Z.-Y. and {Gonzalez-Nuevo}, J. and {B{\'e}thermin}, M. and {Danese}, L.},
        title = "{The Main Sequences of Star-forming Galaxies and Active Galactic Nuclei at High Redshift}",
      journal = {\apj},
         year = 2016,
        month = dec,
       volume = {833},
       number = {2},
          eid = {152},
        pages = {152},
          doi = {10.3847/1538-4357/833/2/152},
archivePrefix = {arXiv},
       eprint = {1610.05910},
 primaryClass = {astro-ph.GA},
       adsurl = {https://ui.adsabs.harvard.edu/abs/2016ApJ...833..152M}
}

@ARTICLE{Bethermin2012,
       author = {{B{\'e}thermin}, Matthieu and {Daddi}, Emanuele and {Magdis}, Georgios and {Sargent}, Mark T. and {Hezaveh}, Yashar and {Elbaz}, David and {Le Borgne}, Damien and {Mullaney}, James and {Pannella}, Maurilio and {Buat}, V{\'e}ronique and {Charmandaris}, Vassilis and {Lagache}, Guilaine and {Scott}, Douglas},
        title = "{A Unified Empirical Model for Infrared Galaxy Counts Based on the Observed Physical Evolution of Distant Galaxies}",
      journal = {\apjl},
         year = 2012,
        month = oct,
       volume = {757},
       number = {2},
          eid = {L23},
        pages = {L23},
          doi = {10.1088/2041-8205/757/2/L23},
archivePrefix = {arXiv},
       eprint = {1208.6512},
 primaryClass = {astro-ph.CO},
       adsurl = {https://ui.adsabs.harvard.edu/abs/2012ApJ...757L..23B}
}

@ARTICLE{Sargent2012,
       author = {{Sargent}, M.~T. and {B{\'e}thermin}, M. and {Daddi}, E. and {Elbaz}, D.},
        title = "{The Contribution of Starbursts and Normal Galaxies to Infrared Luminosity Functions at z < 2}",
      journal = {\apjl},
         year = 2012,
        month = mar,
       volume = {747},
       number = {2},
          eid = {L31},
        pages = {L31},
          doi = {10.1088/2041-8205/747/2/L31},
archivePrefix = {arXiv},
       eprint = {1202.0290},
 primaryClass = {astro-ph.CO},
       adsurl = {https://ui.adsabs.harvard.edu/abs/2012ApJ...747L..31S}
}

@ARTICLE{Ilbert2015,
       author = {{Ilbert}, O. and {Arnouts}, S. and {Le Floc'h}, E. and {Aussel}, H. and {Bethermin}, M. and {Capak}, P. and {Hsieh}, B.-C. and {Kajisawa}, M. and {Karim}, A. and {Le F{\`e}vre}, O. and {Lee}, N. and {Lilly}, S. and {McCracken}, H.~J. and {Michel-Dansac}, L. and {Moutard}, T. and {Renzini}, M.~A. and {Salvato}, M. and {Sanders}, D.~B. and {Scoville}, N. and {Sheth}, K. and {Silverman}, J.~D. and {Smol{\v{c}}i{\'c}}, V. and {Taniguchi}, Y. and {Tresse}, L.},
        title = "{Evolution of the specific star formation rate function at z< 1.4 Dissecting the mass-SFR plane in COSMOS and GOODS}",
      journal = {\aap},
         year = 2015,
        month = jul,
       volume = {579},
          eid = {A2},
        pages = {A2},
          doi = {10.1051/0004-6361/201425176},
archivePrefix = {arXiv},
       eprint = {1410.4875},
 primaryClass = {astro-ph.GA},
       adsurl = {https://ui.adsabs.harvard.edu/abs/2015A&A...579A...2I}
}

@ARTICLE{Rinaldi2022,
       author = {{Rinaldi}, Pierluigi and {Caputi}, Karina I. and {van Mierlo}, Sophie E. and {Ashby}, Matthew L.~N. and {Caminha}, Gabriel B. and {Iani}, Edoardo},
        title = "{The Galaxy Starburst/Main-sequence Bimodality over Five Decades in Stellar Mass at z ≍ 3-6.5}",
      journal = {\apj},
         year = 2022,
        month = may,
       volume = {930},
       number = {2},
          eid = {128},
        pages = {128},
          doi = {10.3847/1538-4357/ac5d39},
archivePrefix = {arXiv},
       eprint = {2112.03935},
 primaryClass = {astro-ph.GA},
       adsurl = {https://ui.adsabs.harvard.edu/abs/2022ApJ...930..128R}
}

@ARTICLE{Rinaldi2025,
       author = {{Rinaldi}, P. and {Navarro-Carrera}, R. and {Caputi}, K.~I. and {Iani}, E. and {{\"O}stlin}, G. and {Colina}, L. and {Alberts}, S. and {{\'A}lvarez-M{\'a}rquez}, J. and {Annunziatella}, M. and {Boogaard}, L. and {Costantin}, L. and {Hjorth}, J. and {Langeroodi}, D. and {Melinder}, J. and {Moutard}, T. and {Walter}, F.},
        title = "{The Emergence of the Star Formation Main Sequence with Redshift Unfolded by JWST}",
      journal = {\apj},
         year = 2025,
        month = mar,
       volume = {981},
       number = {2},
          eid = {161},
        pages = {161},
          doi = {10.3847/1538-4357/adb309},
archivePrefix = {arXiv},
       eprint = {2406.13554},
 primaryClass = {astro-ph.GA},
       adsurl = {https://ui.adsabs.harvard.edu/abs/2025ApJ...981..161R}
}

@ARTICLE{Lara-Lopez2010,
       author = {{Lara-L{\'o}pez}, M.~A. and {Cepa}, J. and {Bongiovanni}, A. and {P{\'e}rez Garc{\'\i}a}, A.~M. and {Ederoclite}, A. and {Casta{\~n}eda}, H. and {Fern{\'a}ndez Lorenzo}, M. and {Povi{\'c}}, M. and {S{\'a}nchez-Portal}, M.},
        title = "{A fundamental plane for field star-forming galaxies}",
      journal = {\aap},
         year = 2010,
        month = oct,
       volume = {521},
          eid = {L53},
        pages = {L53},
          doi = {10.1051/0004-6361/201014803},
archivePrefix = {arXiv},
       eprint = {1005.0509},
 primaryClass = {astro-ph.CO},
       adsurl = {https://ui.adsabs.harvard.edu/abs/2010A&A...521L..53L}
}

@ARTICLE{Mannucci2010,
       author = {{Mannucci}, F. and {Cresci}, G. and {Maiolino}, R. and {Marconi}, A. and {Gnerucci}, A.},
        title = "{A fundamental relation between mass, star formation rate and metallicity in local and high-redshift galaxies}",
      journal = {\mnras},
         year = 2010,
        month = nov,
       volume = {408},
       number = {4},
        pages = {2115-2127},
          doi = {10.1111/j.1365-2966.2010.17291.x},
archivePrefix = {arXiv},
       eprint = {1005.0006},
 primaryClass = {astro-ph.CO},
       adsurl = {https://ui.adsabs.harvard.edu/abs/2010MNRAS.408.2115M}
}

@ARTICLE{Mannucci2011,
       author = {{Mannucci}, F. and {Salvaterra}, R. and {Campisi}, M.~A.},
        title = "{The metallicity of the long GRB hosts and the fundamental metallicity relation of low-mass galaxies}",
      journal = {\mnras},
         year = 2011,
        month = jun,
       volume = {414},
       number = {2},
        pages = {1263-1268},
          doi = {10.1111/j.1365-2966.2011.18459.x},
archivePrefix = {arXiv},
       eprint = {1011.4506},
 primaryClass = {astro-ph.CO},
       adsurl = {https://ui.adsabs.harvard.edu/abs/2011MNRAS.414.1263M}
}

@ARTICLE{Hunt2016,
       author = {{Hunt}, Leslie and {Dayal}, Pratika and {Magrini}, Laura and {Ferrara}, Andrea},
        title = "{Coevolution of metallicity and star formation in galaxies to z ≃ 3.7 - I. A Fundamental Plane}",
      journal = {\mnras},
         year = 2016,
        month = dec,
       volume = {463},
       number = {2},
        pages = {2002-2019},
          doi = {10.1093/mnras/stw1993},
archivePrefix = {arXiv},
       eprint = {1608.05417},
 primaryClass = {astro-ph.GA},
       adsurl = {https://ui.adsabs.harvard.edu/abs/2016MNRAS.463.2002H}
}

@ARTICLE{Zahid2014b,
       author = {{Zahid}, H. Jabran and {Dima}, Gabriel I. and {Kudritzki}, Rolf-Peter and {Kewley}, Lisa J. and {Geller}, Margaret J. and {Hwang}, Ho Seong and {Silverman}, John D. and {Kashino}, Daichi},
        title = "{The Universal Relation of Galactic Chemical Evolution: The Origin of the Mass-Metallicity Relation}",
      journal = {\apj},
         year = 2014,
        month = aug,
       volume = {791},
       number = {2},
          eid = {130},
        pages = {130},
          doi = {10.1088/0004-637X/791/2/130},
archivePrefix = {arXiv},
       eprint = {1404.7526},
 primaryClass = {astro-ph.GA},
       adsurl = {https://ui.adsabs.harvard.edu/abs/2014ApJ...791..130Z}
}

@ARTICLE{Cresci2019,
       author = {{Cresci}, G. and {Mannucci}, F. and {Curti}, M.},
        title = "{Fundamental metallicity relation in CALIFA, SDSS-IV MaNGA, and high-z galaxies}",
      journal = {\aap},
         year = 2019,
        month = jul,
       volume = {627},
          eid = {A42},
        pages = {A42},
          doi = {10.1051/0004-6361/201834637},
archivePrefix = {arXiv},
       eprint = {1811.06015},
 primaryClass = {astro-ph.GA},
       adsurl = {https://ui.adsabs.harvard.edu/abs/2019A&A...627A..42C}
}

@ARTICLE{Curti2020,
       author = {{Curti}, Mirko and {Mannucci}, Filippo and {Cresci}, Giovanni and {Maiolino}, Roberto},
        title = "{The mass-metallicity and the fundamental metallicity relation revisited on a fully T$_{e}$-based abundance scale for galaxies}",
      journal = {\mnras},
         year = 2020,
        month = jan,
       volume = {491},
       number = {1},
        pages = {944-964},
          doi = {10.1093/mnras/stz2910},
archivePrefix = {arXiv},
       eprint = {1910.00597},
 primaryClass = {astro-ph.GA},
       adsurl = {https://ui.adsabs.harvard.edu/abs/2020MNRAS.491..944C}
}

@ARTICLE{Curti2023,
       author = {{Curti}, Mirko and {D'Eugenio}, Francesco and {Carniani}, Stefano and {Maiolino}, Roberto and {Sandles}, Lester and {Witstok}, Joris and {Baker}, William M. and {Bennett}, Jake S. and {Piotrowska}, Joanna M. and {Tacchella}, Sandro and {Charlot}, Stephane and {Nakajima}, Kimihiko and {Maheson}, Gabriel and {Mannucci}, Filippo and {Amiri}, Amirnezam and {Arribas}, Santiago and {Belfiore}, Francesco and {Bonaventura}, Nina R. and {Bunker}, Andrew J. and {Chevallard}, Jacopo and {Cresci}, Giovanni and {Curtis-Lake}, Emma and {Hayden-Pawson}, Connor and {Jones}, Gareth C. and {Kumari}, Nimisha and {Laseter}, Isaac and {Looser}, Tobias J. and {Marconi}, Alessandro and {Maseda}, Michael V. and {Scholtz}, Jan and {Smit}, Renske and {{\"U}bler}, Hannah and {Wallace}, Imaan E.~B.},
        title = "{The chemical enrichment in the early Universe as probed by JWST via direct metallicity measurements at z {\ensuremath{\sim}} 8}",
      journal = {\mnras},
         year = 2023,
        month = jan,
       volume = {518},
       number = {1},
        pages = {425-438},
          doi = {10.1093/mnras/stac2737},
archivePrefix = {arXiv},
       eprint = {2207.12375},
 primaryClass = {astro-ph.GA},
       adsurl = {https://ui.adsabs.harvard.edu/abs/2023MNRAS.518..425C}
}

@ARTICLE{Nakajima2023,
       author = {{Nakajima}, Kimihiko and {Ouchi}, Masami and {Isobe}, Yuki and {Harikane}, Yuichi and {Zhang}, Yechi and {Ono}, Yoshiaki and {Umeda}, Hiroya and {Oguri}, Masamune},
        title = "{JWST Census for the Mass-Metallicity Star Formation Relations at z = 4-10 with Self-consistent Flux Calibration and Proper Metallicity Calibrators}",
      journal = {\apjs},
         year = 2023,
        month = dec,
       volume = {269},
       number = {2},
          eid = {33},
        pages = {33},
          doi = {10.3847/1538-4365/acd556},
archivePrefix = {arXiv},
       eprint = {2301.12825},
 primaryClass = {astro-ph.GA},
       adsurl = {https://ui.adsabs.harvard.edu/abs/2023ApJS..269...33N}
}

@ARTICLE{Salim2014,
       author = {{Salim}, Samir and {Lee}, Janice C. and {Ly}, Chun and {Brinchmann}, Jarle and {Dav{\'e}}, Romeel and {Dickinson}, Mark and {Salzer}, John J. and {Charlot}, St{\'e}phane},
        title = "{A Critical Look at the Mass-Metallicity-Star Formation Rate Relation in the Local Universe. I. An Improved Analysis Framework and Confounding Systematics}",
      journal = {\apj},
         year = 2014,
        month = dec,
       volume = {797},
       number = {2},
          eid = {126},
        pages = {126},
          doi = {10.1088/0004-637X/797/2/126},
archivePrefix = {arXiv},
       eprint = {1411.7391},
 primaryClass = {astro-ph.GA},
       adsurl = {https://ui.adsabs.harvard.edu/abs/2014ApJ...797..126S}
}

@ARTICLE{Torrey2018,
       author = {{Torrey}, Paul and {Vogelsberger}, Mark and {Hernquist}, Lars and {McKinnon}, Ryan and {Marinacci}, Federico and {Simcoe}, Robert A. and {Springel}, Volker and {Pillepich}, Annalisa and {Naiman}, Jill and {Pakmor}, R{\"u}diger and {Weinberger}, Rainer and {Nelson}, Dylan and {Genel}, Shy},
        title = "{Similar star formation rate and metallicity variability time-scales drive the fundamental metallicity relation}",
      journal = {\mnras},
         year = 2018,
        month = jun,
       volume = {477},
       number = {1},
        pages = {L16-L20},
          doi = {10.1093/mnrasl/sly031},
archivePrefix = {arXiv},
       eprint = {1711.11039},
 primaryClass = {astro-ph.GA},
       adsurl = {https://ui.adsabs.harvard.edu/abs/2018MNRAS.477L..16T}
}

@ARTICLE{Sanders2018,
       author = {{Sanders}, Ryan L. and {Shapley}, Alice E. and {Kriek}, Mariska and {Freeman}, William R. and {Reddy}, Naveen A. and {Siana}, Brian and {Coil}, Alison L. and {Mobasher}, Bahram and {Dav{\'e}}, Romeel and {Shivaei}, Irene and {Azadi}, Mojegan and {Price}, Sedona H. and {Leung}, Gene and {Fetherolf}, Tara and {de Groot}, Laura and {Zick}, Tom and {Fornasini}, Francesca M. and {Barro}, Guillermo},
        title = "{The MOSDEF Survey: A Stellar Mass-SFR-Metallicity Relation Exists at z {\ensuremath{\sim}} 2.3}",
      journal = {\apj},
         year = 2018,
        month = may,
       volume = {858},
       number = {2},
          eid = {99},
        pages = {99},
          doi = {10.3847/1538-4357/aabcbd},
archivePrefix = {arXiv},
       eprint = {1711.00224},
 primaryClass = {astro-ph.GA},
       adsurl = {https://ui.adsabs.harvard.edu/abs/2018ApJ...858...99S}
}

@ARTICLE{Pettini2004,
       author = {{Pettini}, Max and {Pagel}, Bernard E.~J.},
        title = "{[OIII]/[NII] as an abundance indicator at high redshift}",
      journal = {\mnras},
         year = 2004,
        month = mar,
       volume = {348},
       number = {3},
        pages = {L59-L63},
          doi = {10.1111/j.1365-2966.2004.07591.x},
archivePrefix = {arXiv},
       eprint = {astro-ph/0401128},
 primaryClass = {astro-ph},
       adsurl = {https://ui.adsabs.harvard.edu/abs/2004MNRAS.348L..59P}
}

@ARTICLE{Kewley2002,
       author = {{Kewley}, L.~J. and {Dopita}, M.~A.},
        title = "{Using Strong Lines to Estimate Abundances in Extragalactic H II Regions and Starburst Galaxies}",
      journal = {\apjs},
         year = 2002,
        month = sep,
       volume = {142},
       number = {1},
        pages = {35-52},
          doi = {10.1086/341326},
archivePrefix = {arXiv},
       eprint = {astro-ph/0206495},
 primaryClass = {astro-ph},
       adsurl = {https://ui.adsabs.harvard.edu/abs/2002ApJS..142...35K}
}

@ARTICLE{Tremonti2004,
       author = {{Tremonti}, Christy A. and {Heckman}, Timothy M. and {Kauffmann}, Guinevere and {Brinchmann}, Jarle and {Charlot}, St{\'e}phane and {White}, Simon D.~M. and {Seibert}, Mark and {Peng}, Eric W. and {Schlegel}, David J. and {Uomoto}, Alan and {Fukugita}, Masataka and {Brinkmann}, Jon},
        title = "{The Origin of the Mass-Metallicity Relation: Insights from 53,000 Star-forming Galaxies in the Sloan Digital Sky Survey}",
      journal = {\apj},
         year = 2004,
        month = oct,
       volume = {613},
       number = {2},
        pages = {898-913},
          doi = {10.1086/423264},
archivePrefix = {arXiv},
       eprint = {astro-ph/0405537},
 primaryClass = {astro-ph},
       adsurl = {https://ui.adsabs.harvard.edu/abs/2004ApJ...613..898T}
}
\endgroup

\begin{appendix}
\section{Choice of the metallicity parameters}\label{sec:appendix}
In Section \ref{sec:metallicity} we introduced the FMR functional form and we explained how we can compute its parameters from the local MZR and main sequence. We also introduced the three free metallicity parameters used to compute the others: $\nabla_{\textrm{FMR},0}$, $a_\textrm{MZR}$, and $Z_{\textrm{O}/\textrm{H},0}$ and we provided the values used for these parameters in Table \ref{table:parameters}. In this Appendix, we motivate this choice, showing that those values encompass most of the values found in literature.

The slope of the local MZR, $a_{\textrm{MZR}}$ has been studied by several authors \citep{Kewley2002, Tremonti2004, Maiolino2008, Mannucci2009, Andrews2013, Zahid2014b, Curti2020, Sanders2021}. Most recent studies \citep{Curti2020, Sanders2021} favor shallower values $\sim 0.3$ than those found in previous ones $\sim 0.6$ \citep{Andrews2013, Zahid2014b}. These values are found by fitting the metallicity as a function of the stellar mass irrespectively of the galaxy SFR. $a_{\textrm{MZR}}$ in the range $0.3-0.6$ is adopted even in \citetalias{Chruslinska2021} for their explorations. However, we notice that most of the FMR reported in literature (e.g., \citealt{Andrews2013, Hunt2016, Curti2020}) imply even shallower values of $a_{\textrm{MZR}}$ when coupled with the main sequence \cite{Popesso2023}. This is because \cite{Popesso2023} main sequence features a steep faint-end slope $a_{\textrm{MS}}\sim1$. This means that to an increase in $M_\star$ corresponds a relatively large SFR increase, thus weakening the metallicity dependence on $M_\star$. For example, the FMR of \cite{Andrews2013} features $\gamma=0.43$, $\nabla_{\textrm{FMR},0}\sim 0.27$ and that of \cite{Curti2020} $\gamma=0.31$, $\nabla_{\textrm{FMR},0}\sim0.17$, corresponding to $a_{\textrm{MZR}}\sim 0.15$ in the case $a_{\textrm{MS}}=1$\footnote{Note that these two works present also determinations of the MZR by stacking together galaxies with the same stellar mass and different SFR. They find higher $a_\textrm{MZR}$ values ($\sim0.3$ for \cite{Andrews2013}, and $\sim0.6$ for \cite{Curti2020}). The difference with respect to $a_\textrm{MZR}\sim 0.15$ is due to the galaxy sample chosen, which feature a much shallower dependence of SFR on $M_\star$ than the \cite{Popesso2023} main sequence.}. To bracket all the possible range of uncertainty, in this work, we show results for $a_{\textrm{MZR}}=0.15$, 0.3, 0.6.

$\nabla_{\textrm{FMR},0}$ represents the strength of the anti-correlation between metallicity and SFR. Several works estimate its value but the results are dependent on metallicity calibration and/or redshift \citep{Salim2014, Sanders2018, Torrey2018, Sanders2021}. Most of these works report values ranging between $0.2-0.3$, with some exception for specific mass or redshift bins arriving at $\nabla_{\textrm{FMR},0}\gtrsim 0.1$ (see \citetalias{Chruslinska2021} for a thorough discussion). However, \citetalias{Chruslinska2021} find that values $\nabla_{\textrm{FMR},0}\lesssim0.17$ underpredict the redshift evolution of the MZR at $z\lesssim1.5$ and the scatter of the MZR. For these reasons, here we show results for $\nabla_{\textrm{FMR},0}=0.2$, 0.3.

We also vary the overall normalization of the FMR $Z_{\textrm{O}/\textrm{H},0}$. Direct method measurements of gas phase metallicity \citep{Pettini2004, Curti2020, Sanders2021} derive $Z_{\textrm{O}/\textrm{H},0}\sim 8.8$, which is lower with respect to the values found using theoretical calibrations $Z_{\textrm{O}/\textrm{H},0}\sim 9.1$ \citep{Kewley2002, Tremonti2004, Kobulnicky2004, Mannucci2010}. This reflects a longstanding issue where direct methods tend to underestimate oxygen abundance, while theoretical models tend to overestimate it \citep{Maiolino2019}. \cite{Chruslinska2024, Chruslinska2025} show that, in order to reconcile the average [O/Fe] of young star-forming galaxies and old Milky Way stars, a shift of $0.2$ above the direct method estimates might be necessary. In this work we show results for $Z_{\textrm{O}/\textrm{H},0}=8.8,9.0,9.2$, so to encompass all the possible values of $Z_{\textrm{O}/\textrm{H},0}$.
\end{appendix}

\end{document}